\documentclass[reprint,aps,showpacs,prx,superscriptaddress,floatfix,10pt,nofootinbib,longbibliography]{revtex4-2}
\usepackage[T1]{fontenc}
\usepackage[latin9]{inputenc}
\usepackage{amsmath}
\usepackage{amssymb}
\usepackage{esint}
\usepackage{graphicx}
\usepackage{bbold}
\usepackage{appendix}
\usepackage{verbatim}
\usepackage{bm}
\usepackage{xr-hyper}

\makeatletter

\usepackage{color}
     
\usepackage{physics}  
\usepackage{xcolor} 

\definecolor{mygreen}{RGB}{0, 126, 0}
\definecolor{myorange}{RGB}{255, 136, 19}
\definecolor{mymagenta}{RGB}{200, 0, 100}

\def\@fnsymbol#1{\ensuremath{\ifcase#1\or \dagger\or \ddagger\or
   \mathsection\or \mathparagraph\or \|\or **\or \dagger\dagger
   \or \ddagger\ddagger \else\@ctrerr\fi}}
    \makeatother

\makeatother

\begin{document}
\title{Simulating one-dimensional quantum chromodynamics on a quantum computer: Real-time evolutions of tetra- and pentaquarks}

\author{Yasar Y. Atas$^*$}
\email{yasar.yilmaz.atas@gmail.com}
\affiliation{Institute for Quantum Computing, University of Waterloo, Waterloo, ON, Canada, N2L 3G1}
\affiliation{Department of Physics \& Astronomy, University of Waterloo, Waterloo, ON, Canada, N2L 3G1}
\author{Jan F. Haase$^*$}
\email{jan.frhaase@gmail.com}
\affiliation{Institute for Quantum Computing, University of Waterloo, Waterloo, ON, Canada, N2L 3G1}
\affiliation{Department of Physics \& Astronomy, University of Waterloo, Waterloo, ON, Canada, N2L 3G1}
\affiliation{Institut f\"ur Theoretische Physik und IQST, Universit\"at Ulm, Albert-Einstein-Allee 11, D-89069 Ulm, Germany}
\author{Jinglei Zhang}
\email{jingleizl@gmail.com}
\affiliation{Institute for Quantum Computing, University of Waterloo, Waterloo, ON, Canada, N2L 3G1}
\affiliation{Department of Physics \& Astronomy, University of Waterloo, Waterloo, ON, Canada, N2L 3G1}

\author{Victor Wei}
\affiliation{Institute for Quantum Computing, University of Waterloo, Waterloo, ON, Canada, N2L 3G1}
\affiliation{Department of Physics, McGill University, Montreal, QC, Canada, H3A 2T8}

\author{Sieglinde M.-L. Pfaendler}
\affiliation{IBM Deutschland Research \& Development GmbH, Sch\"onaicher Str. 220, D-71032 B\"oblingen, Germany}

\author{Randy Lewis}
\affiliation{Department of Physics and Astronomy, York University, Toronto, ON, Canada, M3J 1P3}

\author{Christine A. Muschik}
\affiliation{Institute for Quantum Computing, University of Waterloo, Waterloo, ON, Canada, N2L 3G1}
\affiliation{Department of Physics \& Astronomy, University of Waterloo, Waterloo, ON, Canada, N2L 3G1}
\affiliation{Perimeter Institute for Theoretical Physics, Waterloo, ON, Canada, N2L 2Y5}
\date{\today}

\def\thefootnote{*}\footnotetext{These authors contributed equally to this work}

\begin{abstract}
Quantum chromodynamics - the theory of quarks and gluons - has been known for decades, but it is yet to be fully understood. A recent example is the prediction and experimental discovery of tetraquarks, that opened a new research field.
Crucially, numerous unsolved questions of the standard model can exclusively be addressed by nonperturbative calculations. 
Quantum computers can solve problems for which 
well established QCD methods are inapplicable, such as real-time evolution. We take a key step in exploring this possibility by performing a real-time evolution of tetraquark and pentaquark physics in one-dimensional SU(3) gauge theory on a superconducting quantum computer. Our experiment represents a first quantum computation involving quarks with three colour degrees of freedom, \textit{i.e.} with the gauge group of QCD.
\end{abstract}

\maketitle

\section{Introduction} 
Quantum chromodynamics (QCD) provides the fundamental understanding of the strong nuclear force. It describes a vast range of hadrons and their properties in terms of just the quark masses and a gauge coupling. The recent discoveries~\cite{Chen:2022asf} of several tetraquark candidates are reminders of the richness still remaining to be understood within QCD.

Lattice gauge theory is the first-principles nonperturbative theoretical tool for studying QCD. Emerging quantum computers will allow lattice studies to access new topics within QCD, such as real-time evolution~\cite{banuls_simulating_2020,bauer_quantum_2022,dalmonte2016lattice}. In this work, we use real-time evolution to present the first study of tetraquarks and pentaquarks on a quantum computer. Our calculations use SU(3) gauge theory in one spatial dimension \cite{silvi_tensor_2019}, and the number of qubits (entangling gates) required scales only linearly (quadratically) in the number of lattice sites $N$. To match the available quantum hardware, our experimental demonstrations focus on a lattice of minimal length, \textit{i.e.} the basic building block.

Previous quantum computations within U(1) gauge theory~\cite{martinez_real-time_2016,klco_quantum-classical_2018,kokail_self-verifying_2019,lu_simulations_2019,mil_scalable_2020,surace_lattice_2020,yang_observation_2020,zhou_thermalization_2021,nguyen_digital_2022} showed electron-positron pair production. Moving from this simple Abelian case to a more complex non-Abelian theory reveals qualitatively new phenomena. For example, in addition to quark-antiquark pair production (and the existence of a meson), there is also a gauge-singlet particle having valence quarks without valence antiquarks (\textit{i.e.} the baryon). A recent paper~\cite{atas2021} presented the first quantum computation of a baryon mass in a SU(2) gauge theory.

In the present work we consider SU(3), which is the gauge group of QCD, and demonstrate color neutral objects (also called gauge singlets). Color neutral states of SU(3) are invariant under arbitrary rotations in color space and thus involve all the color components (charges) available in the theory \textit{i.e.} red, green, and blue (and their anticolor counterpart). This is in contrast to Abelian quantum electrodynamics (QED), where a singlet state involves electron-positron pairs only. Since the color singlet states are the relevant physical states, their study and simulation constitute an important step towards the understanding, description and prediction of more complex and realistic experiments.

In order to study the properties and interactions of the gauge singlets, we perform two experiments. First, we perform a quantum simulation of the tetraquark. Specifically, we identify the state possessing two quarks in a colour antitriplet plus two antiquarks in a colour triplet. The mixing of this state with a baryon-antibaryon pair and with other quantum states is extracted from the quantum computation of time evolution. Second, we study a pentaquark by considering two quark flavours of different masses. Taking the infinite-mass limit for the second flavour allows us to perform an experiment showing oscillations between a pentaquark and a baryon.

Very recent quantum simulations of SU(2) and SU(3) gauge theories for particle physics~\cite{klco_su2_2020,ARahman:2021ktn,atas2021,ciavarella_trailhead_2021,Illa:2022jqb,rahman2022,Fromm:2022vaj,farrell_preparations_2022,farrell2022preparations} have succeeded in accessing increasingly complex model systems on the route towards QCD. 
Our experiment takes this quest a crucial step forward by arriving at the simulation of physical states of SU(3) relevant for hadron physics experiments.


\begin{figure}[t]
    \includegraphics[width=\columnwidth]{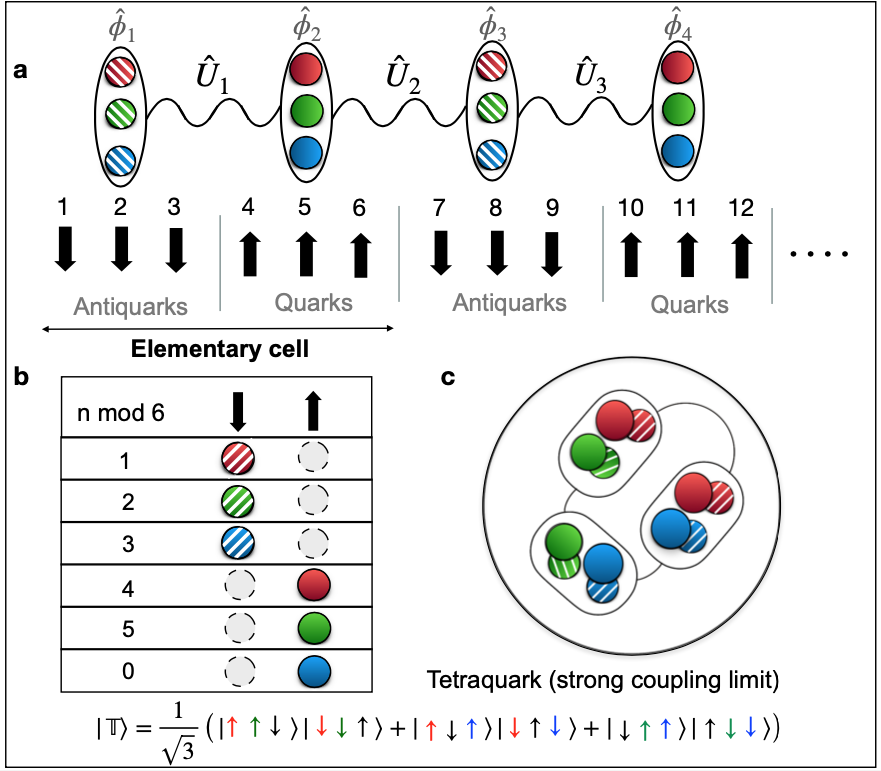} 

\caption{\textbf{Gauge theory on the lattice}. In order to study the SU(3) gauge theory in one dimension, we use the lattice shown in panel \textbf{a}, where space is discretised into unit cells that each hold up to three quarks (red, green, blue) represented by filled circles and up to three antiquarks (antired, antigreen, antiblue) represented by striped circles. Each unit cell is connected to the neighbouring one with a gauge link $\hat{U}_{i}$. Each particle represented by the fermionic field $\hat{\phi}_{n}^{i}$ ($i=1,2,3$) is mapped to a qubit according to the staggerisation table in panel \textbf{b}, where $n$ is the index of the qubit. As an example, panel \textbf{c} shows a pictorial representation of the tetraquark state for an elementary cell in the strong coupling limit, along with its spin representation (see text for details).} \label{fig:tetraquark}
\end{figure}

\section{Theory}
\subsection{SU(3) gauge theory}
Our calculations use the Hamiltonian approach where time is not discretized, and the lattice is purely spatial. We consider a one-dimensional (1D) lattice with open boundary conditions, where each site $n$ hosts a fermionic field with three colour components, $\boldsymbol{\hat{\phi}}_{n}=\left(\hat{\phi}_{n}^{1}, \hat{\phi}_{n}^{2}, \hat{\phi}_{n}^{3}\right)^{\mathrm{T}} $.  We choose to work with staggered fermions~\cite{kogut_hamiltonian_1975} with the convention that odd sites host antimatter, while even sites host matter, as shown in Fig.~\ref{fig:tetraquark}a. The gauge fields are defined on the link between sites $n$ and $n+1$, and mediate the interaction between colour degrees of freedom.
The gauge-invariant lattice Hamiltonian in natural units ($\hbar=c=1$) reads 
\begin{align}
\hat{H}_l = \notag & \frac{1}{2a} \sum_{n=1}^{N-1} \left( \hat{\phi}_{n}^{\dagger} \hat{U}_{n} \hat{\phi}_{n+1} + \operatorname{H.c.}\right) \\ 
&+ m \sum_{n=1}^{N} (-1)^{n} \hat{\phi}_{n}^{\dagger} \hat{\phi}_{n} + \frac{a g^{2}}{2} \sum_{n=1}^{N-1} \mathbf{\hat{L}}_{n}^{2} \label{KSham},
\end{align}
where $\operatorname{H.c.}$  denotes the Hermitian conjugate, $N$ is the number of
lattice sites with spacing $a$, $m$ is the quark bare mass, and $g$ is the bare coupling. 
The first term in the  Hamiltonian describes the creation of particle-antiparticle pairs with $\hat{U}_{n}$ the corresponding gauge operator adapting the gauge field during pair creation. The second term is the mass term (the alternating sign appearing here is the signature of the staggered formulation).  The last term  encodes the colour electric energy of the system and is expressed in terms of the left colour electric field $\hat{\mathbf{L}}_{n}$ on the link $n$. Furthermore, it is convenient to introduce the non-Abelian charges at site $n$,  $\hat{Q}_{n}^{a}=\sum_{i,j=1}^{3}\hat{\phi}_{n}^{i\dagger}(T^{a})_{ij}\hat{\phi}_{n}^{j}$
where $T^{a}=\lambda^{a}/2$, and $\lambda^{a}$  ($a=1,\dots,8$) are the Gell-Mann matrices~\cite{griffiths2020introduction}. These charges appear in the non-Abelian version of the Gauss law that physical states must satisfy~\cite{zohar_quantum_2015}.
We work in the sector with zero external charges and zero total non-Abelian charge, \textit{i.e.} a colour singlet state must satisfy $\hat{Q}_{tot}^{a}\ket{\Psi}\equiv\sum_{n}\hat{Q}_{n}^{a}\ket{\Psi}=0$. Besides the 
eight non-Abelian charges, the Hamiltonian also conserves the baryon number $B$, which measures the  matter-antimatter imbalance (see App.~\ref{app:qubit_formulation}). In our first study, we target tetraquark physics and are therefore interested in the $B=0$ subsector where all states contain an equal number of quarks and antiquarks. In our second study we target pentaquark physics and consider therefore the sector with $B=1$.

\subsection{Effective qubit formulation}
To simulate and study the rich physics of the SU(3) theory, we encode Eq.~(\ref{KSham}) in a Hamiltonian suitable for quantum simulations.
In a first step, a gauge transformation is applied to eliminate the gauge degrees of freedom from the Hamiltonian, allowing us to express the Hamiltonian in terms of fermions only~\cite{kuhn_non-abelian_2015}. This first step is applied to save resources (as gauge fields are not stored explicitly in the qubit register) at the expense of introducing long-range interactions. In a second step, a Jordan-Wigner transformation~\cite{JordanWigner} translates fermionic matter degrees of freedom into spin $\frac{1}{2}$, \textit{i.e.} qubit degrees of freedom (see Fig.~\ref{fig:tetraquark}). 
It is convenient to rescale the Hamiltonian with the lattice spacing $a$, resulting in
\begin{equation}\label{Eq_Full_Encoded_Hamiltonian}
\hat{{H}}=\hat{H}_{kin}+\tilde{m}\hat{H}_{m}+\frac{1}{2x}\hat{H}_{e},
\end{equation}
where $\tilde{m}=am$  and $x=1/(ga)^2$ are the dimensionless mass and coupling constant respectively. 
In the spin formulation, the kinetic term is given by
\begin{align}
    \hat{H}_{kin}=\notag \frac{1}{2}\sum_{n=1}^{N-1}(-1)^n& \left(\hat{\sigma}_{3n-2}^{+}\hat{\sigma}_{3n-1}^{z}\hat{\sigma}_{3n}^{z} \hat{\sigma}_{3n+1}^{-}\right. \\
   &\notag -\left. \hat{\sigma}_{3n-1}^{+}\hat{\sigma}_{3n}^{z}\hat{\sigma}_{3n+1}^{z} \hat{\sigma}_{3n+2}^{-}\right. \\
   & +\left. \hat{\sigma}_{3n}^{+}\hat{\sigma}_{3n+1}^{z}\hat{\sigma}_{3n+2}^{z} \hat{\sigma}_{3n+3}^{-} +\operatorname{H.c.}\right), \label{kinetic_ham_qubit}
\end{align}
and the mass term reads
\begin{equation}
    \hat{H}_{m}=\frac{1}{2}\sum_{n=1}^{N}\left[(-1)^{n}\left(\hat{\sigma}_{3n-2}^{z}+\hat{\sigma}_{3n-1}^{z}+\hat{\sigma}_{3n}^{z}\right)+3\right]. \label{mass_ham_qubit}
\end{equation}
The operators 
$\hat{\sigma}^x=(\hat{\sigma}^- +\hat{\sigma}^+)$, $\hat{\sigma}^y=i(\hat{\sigma}^- -\hat{\sigma}^+)$ and $\hat{\sigma}^{z}$ are the usual Pauli matrices.
The colour electric field Hamiltonian takes the form
\begin{equation}
    \hat{H}_{e}=\sum_{n=1}^{N-1}\left( \sum_{m\leq n}\mathbf{\hat{Q}}_{m}\right)^2, \label{electric_ham_qubit}
\end{equation}
where $\hat{\mathbf{Q}}_{m}$ is a vector with eight components given by the non-Abelian charges at site $m$.
The expression of the non-Abelian charges in terms of qubit operators can be found in App.~\ref{app:qubit_formulation}. 
In Sec.~\ref{subsec:GaugeTimeEvo} we use this equation to simulate the time evolution of the electric field energy.

\begin{figure}[t]
   \includegraphics[width=\columnwidth]{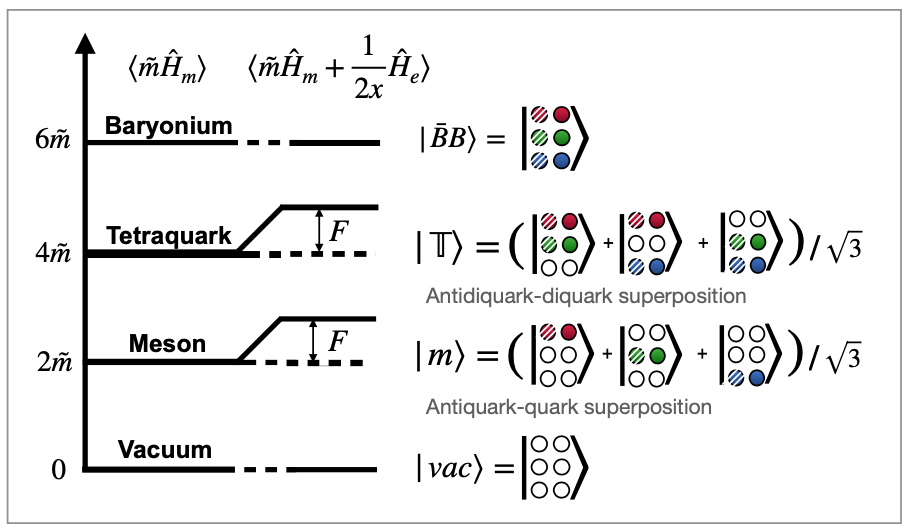} 
\caption{\textbf{Strong coupling states}. The energy eigenstates in the strong coupling limit form a convenient basis for $N=2$ lattice sites (see text for more details). For the basic building block of the lattice (see Fig.~\ref{fig:tetraquark}), these are given by the bare vacuum $|vac\rangle$, meson $|m\rangle$, tetraquark $|\mathbb{T}\rangle$, and baryon-antibaryon (baryonium) $|\bar{B}B\rangle$. The meson and tetraquark both carry one unit of electric flux $F=\frac{4}{3}\times \frac{1}{2x}$ for SU(3), this is indicated by the energy splitting in presence of the electric field term $\hat{H}_{e}$. We resort to a two column representation for the states in the fermion occupation number basis, where the first and second column indicates the state of the antimatter and matter respectively.}\label{fig:strong_coupling_basis}
\end{figure}
\begin{figure*}[t]
    \includegraphics[width=1\textwidth]{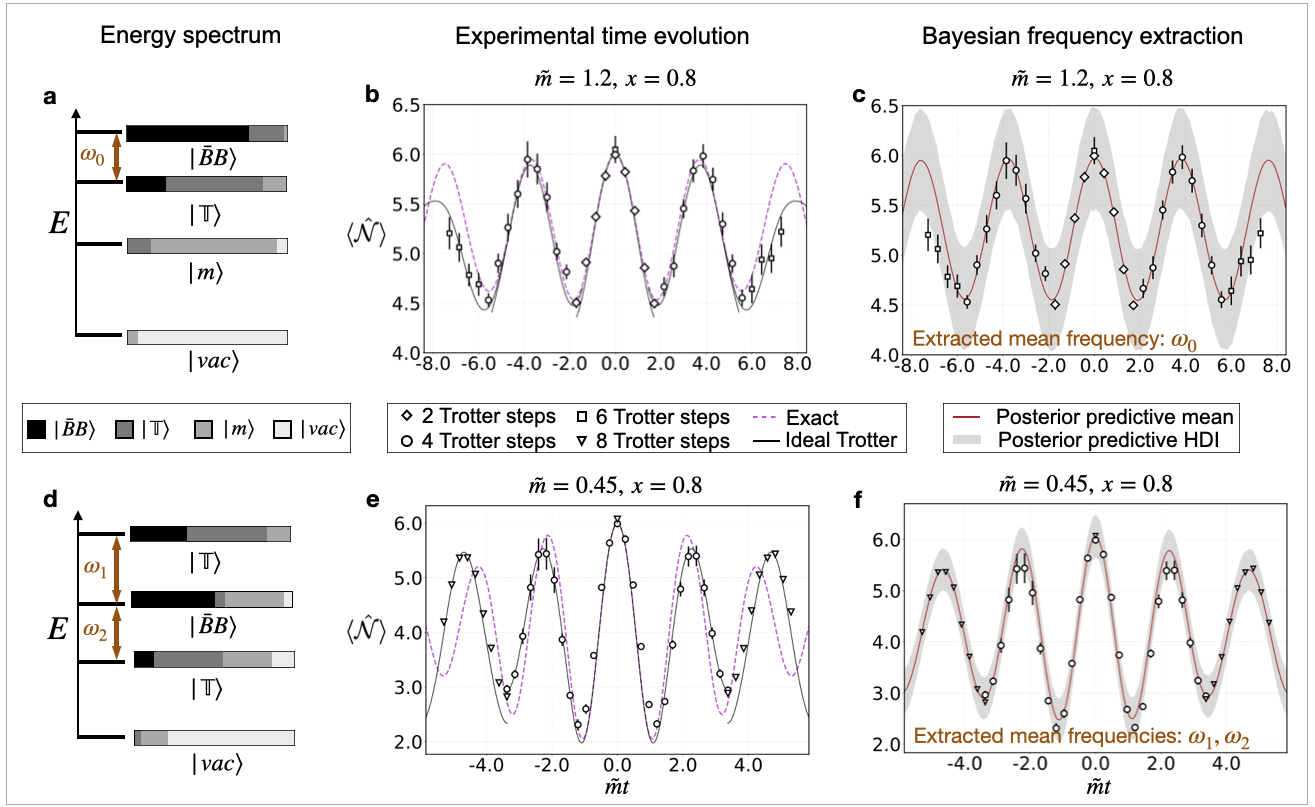} 
\caption{\textbf{Trotter time evolution with one flavour.} We perform experiments for the Hamiltonian parameters $\tilde{m}=1.2$, $x=0.8$ (panels \textbf{a,b,c}) and $\tilde{m}=0.45$, $x=0.8$ (panels \textbf{d,e,f}). The energy spectra on the left are shown to scale and the bars reflect the proportion to which the strong coupling states in Fig.~\ref{fig:strong_coupling_basis} contribute to the individual energy eigenstates. The data obtained on three different IBM quantum computers (panel \textbf{b}: \texttt{ibm\_peekskill}, panel \textbf{e}: \texttt{ibm\_geneva} for $N_\mathrm{T}=4$ and \texttt{ibmq\_lima} for $N_\mathrm{T}=8$) is shown in the middle column, where the different markers denote a different number of Trotter steps, and hence different circuit lengths. The circuit is heavily optimized (see App.~\ref{app:trotter}) and contains $N_\mathrm{T}\cdot 10$ $CNOT$ gates, and a total circuit depth of $N_\mathrm{T}\cdot 25 + 1$ after transpilation to the employed native gates. For the data points shown in the figure, the error mitigation has already been applied. The dashed lines mark the exact Trotter evolution obtained via a numerical exponentiation. We further composed an expected graph for the evolution obtained from the Trotter protocol, which is plotted in a solid black line. The error bars shown here originate via bootstrapping of the error mitigation method \cite{bootstrap}. Error bars corresponding to the quantum projection noise are small due to the 2048 performed shots and would be hidden by the size of the markers and hence not shown. To obtain the energy differences indicated by the arrows in the left column (panel \textbf{a} and \textbf{d}), we resort to Bayesian inference. The results are shown in panels \textbf{c} and \textbf{f} respectively, where the solid lines denote the mean of 5000 samples drawn from the posterior predictive distribution (see App.~\ref{app:bayes}). From these samples we also compute the highest density interval (HDI) equivalent, \textit{i.e.} the grey area marks the interval between the $2.5$ and $97.5$ percentiles. The point estimate for the energy gap between $|\bar{B}B\rangle$ and $|\mathbb{T}\rangle$ (panel \textbf{a}) is given by $\omega_0 = (2\pi)\cdot0.262/[\tilde{m}t]$ with an HDI of $(2\pi)\cdot[0.254,\,0.267]/[\tilde{m}t])$. For the second case in panel \textbf{d}, we find that $\omega_1 = (2 \pi) \cdot 0.482 / [\tilde{m} t]$ with $\mathrm{HDI} = (2 \pi) \cdot [0.409,\, 0.545]/ [\tilde{m} t]$ and $\omega_2 = (2 \pi) \cdot 0.427/ [\tilde{m} t]$ where $\mathrm{HDI} = (2 \pi) \cdot [0.297,\, 0.502]/ [\tilde{m} t]$. Values for other parameters in our probabilistic model can be found in the Appendix \ref{appendix:error_mitigation}.}\label{fig:results}
\end{figure*}
\section{Quantum simulation and results}

In the following, we describe our quantum simulation approach first for the tetraquark study, and then for the pentaquark. For both, we focus on the basic building block consisting of $N=2$ lattice sites (the scaling analysis for larger lattices is given in App.~\ref{app:scalability}). A convenient basis is the strong coupling one given by $\tilde{m}\to\infty$ and $x\rightarrow0$, \textit{i.e.} in the limit in which $\hat{H}_m$ and $\hat{H}_{e}$ dominate over the kinetic term $\hat{H}_{kin}$. The gauge-invariant (colour neutral) basis states in that limit can be constructed by successively applying the kinetic term to the vacuum state. The different basis states obtained for $B=0$ are depicted in Fig.~\ref{fig:strong_coupling_basis} in the fermion occupation basis, where the first and second column of the ket describe the antimatter and matter content of the state respectively.
While all of them possess the same quantum baryon number, the number of particle-antiparticle pairs contained in them differ. These states are all eigenstates of the mass Hamiltonian given in Eq.~(\ref{mass_ham_qubit}), which counts the number of particles and antiparticles in a state, with integer eigenvalues $0,2,4$ and $6$ for the vacuum, meson,  tetraquark and baryon-antibaryon states respectively. The meson state consists of a colour-singlet superposition of particle-antiparticle pairs. By contrast the tetraquark state is a colour superposition of diquark-antidiquark pairs. Like a single antiquark, the diquark is a colour antitriplet state.
 
 At finite coupling $x$ and mass $\tilde{m}$, the eigenstates of the Hamiltonian are given by a superposition of the strong coupling basis states. By studying time evolution under the Hamiltonian in Eq.~(\ref{Eq_Full_Encoded_Hamiltonian}), we can probe the transitions between the different eigenstates. In particular, by choosing the initial state as the strong coupling limit baryon-antibaryon state (containing six particles and antiparticles in total) and in the regime where $x/\tilde{m}\leq 1$, we can probe a single transition between the baryon-antibaryon and the tetraquark state (see Fig.~\ref{fig:results}a). When the parameters are chosen outside of this regime, more than one transition becomes involved in the time evolution, which makes the dynamics richer and more complex.

The time evolution is obtained from a Trotter decomposition~\cite{lloyd1996} that we optimize for minimal gate depth. Figure~\ref{fig:trotter_circuit} in the Appendix shows the Trotter circuit for a basic building block ($N=2$). While this minimal lattice is described by six spins (compare Fig.~\ref{fig:tetraquark}),
we can simulate the $B=0$ sector using only three qubits, due to the existence of a particle-antiparticle symmetry (see App.~\ref{app:qubit_formulation}).
We are interested in tracking the particle number expectation value
\begin{equation}
    \langle \hat{\mathcal{N}}(t)\rangle=
    \bra{\Psi_{0}}e^{it\hat{H}^{(3)}}\hat{H}^{(3)}_{m}e^{-it\hat{H}^{(3)}}\ket{\Psi_{0}}, \label{time_evolved_particle_number}
\end{equation}
as we evolve the system in time starting from an initial state $\ket{\Psi_{0}}$, with $\hat{H}^{(3)}$ the three-qubits Hamiltonian derived in App.~\ref{app:reduction_to_three_qubits}, and given in Eq.~\eqref{eq:three_qubit_ham}. We focus here on $\ket{\Psi_{0}}=\ket{\bar{B}B}$, the baryon-antibaryon state in the strong coupling limit (see Fig.~\ref{fig:strong_coupling_basis}). In terms of spins, the strong coupling baryon-antibaryon state is given by $\ket{\Psi_{0}}=\ket{\downarrow\downarrow\downarrow }\ket{\uparrow\uparrow\uparrow }$, where the first ket refers to antiquarks and the second to quarks (note that only the first ket is implemented in the quantum simulation and the second is implied, as explained in App.~\ref{app:qubit_formulation}).

For our pentaquark study, we consider two quark flavours, with light and heavy quark masses $\tilde{m} $ and $\tilde{M}$ respectively (see Fig.~\ref{fig:penta_result}a).
This scenario can be treated with full generality with quadratic scaling in the required resources. However, to allow for the observation of pentaquark physics on currently available quantum processors, we further reduce the resource requirements by introducing the following steps.
We assume $\tilde{M} \gg \tilde{m}$, such that the heavy quarks can be treated in the infinite mass limit $\tilde{M}\rightarrow \infty$. In this case, the heavy quarks in the model become stationary, \textit{i.e.} their motional degrees of freedom become static. These ``motionally static quarks'' nevertheless contribute through their dynamic colour degrees of freedom.

As a concrete example, we choose to study pentaquarks with two heavy quarks (see Fig.~\ref{fig:penta_result}b). As explained in detail in App.~\ref{app:trotter_penta}, we derive a modified Hamiltonian for this specific case, and devise a scheme that relegates the parts of the calculation that do not have to be performed quantumly to a classical computer. We also show how to split the resulting time evolution into three separate colour subsectors, which allows us to simulate the problem with only four qubits. 
As described in the next section, we apply this resource-efficient time evolution scheme to the baryon state as initial state and observe oscillations between the baryon $\ket{\mathbb{B}}$ and the pentaquark $|\mathbb{P}\rangle$ (see Fig.~\ref{fig:penta_result}c).

\subsection{Error mitigation}
For both studies, we use the self-mitigation method introduced in Ref.~\cite{rahman2022} (see Appendix \ref{appendix:error_mitigation} for details). 
The basic idea is to use our quantum circuit in two ways.  
A ``physics run'' applies the desired number of Trotter steps, $N_{\mathrm{T}}$, forward in time to reach the final time of interest.  
A ``mitigation run'' applies $N_{\mathrm{T}}/2$ steps forward in time followed by $N_{\mathrm{T}}/2$ steps backward in time, which results in a noisy experimental determination of the known initial state.  
Randomized compiling is used to surround the $CNOT$ gates with Pauli gates that turn coherent errors into incoherent errors. 
We find that the minimum number of physics and mitigation runs can be as low as 40 each, and up to 560 depending on the quantum computing device chosen. 
Throughout this work we always collect 2048 shots from a single circuit execution.  
As described in Ref.~\cite{rahman2022}, the noisy observed outcomes measured during the mitigation runs provide an excellent error mitigation for the physics runs when compared with the true expected values.
As in Ref.~\cite{atas2021}, each separate calculation is further accompanied by a set of $2^3$ calibration circuits ($2^4$ for calculations involving the pentaquark) to estimate the transfer map mixing the true outcome probabilities into the observed ones.
Let us remark that such procedure is not preventing the scalability of our calculations. The increased effort using that technique is mainly influenced by the randomized compiling, which has recently been suggested to be linear in the circuit depth \cite{Hashim2021Randomized}. Further, the addition of the mitigation runs increases the computational effort by a constant factor of two, which results in an overall linear scaling of the mitigation efforts in the curcuit depth.

\subsection{Experiment}
For our tetraquark study, we perform two Trotter time evolution experiments on a universal superconducting quantum computer~\cite{IBMQ} using up to eight Trotter time steps. In both experiments, the system is initialized in the strong coupling baryon-antibaryon state $|\bar{B}B\rangle$ and evolved in time under the gauge-invariant three-qubit Hamiltonian given in App.~\ref{app:reduction_to_three_qubits}. Since the Hamiltonian preserves the baryon number, the observed evolution remains within the $B=0$ sector.

\begin{figure*}[t]
    \includegraphics[width=1\textwidth]{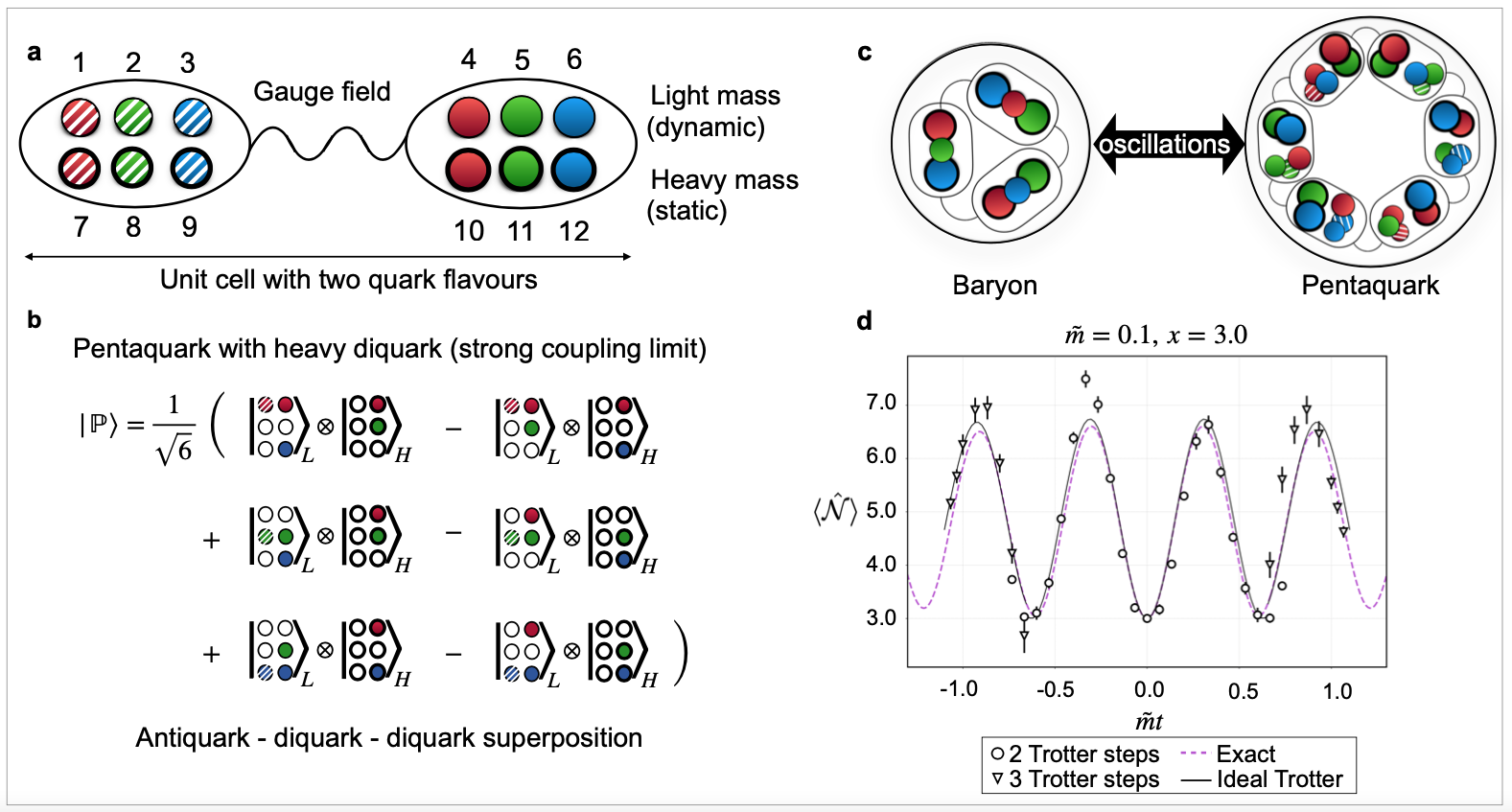} 
\caption{\textbf{Trotter time evolution with two flavours.} We consider two quark flavours, with light and heavy mass respectively. Panel \textbf{a} shows the 12-qubit unit cell for this system with a gauge field connection. Circles with thick rims represent lattice sites hosting heavy quarks (lower row), that can be considered static, in contrast to the dynamic light mass quarks in the upper half of the unit cell. Panel \textbf{b} shows a pentaquark state involving a heavy diquark in the strong coupling limit.  We simulate time evolutions that start with a baryon as initial state and are dominated by baryon - pentaquark oscillations as illustrated in panel \textbf{c}. Panel \textbf{d} shows experimental Trotter time  evolution data for $\tilde{m}=0.1$ and $x=3.0$. The experimental data have been obtained from \texttt{ibmq\_lima} with two and four Trotter steps respectively. Averaging for the error mitigation has been performed over 140 repetitions, where the result of the physical runs as well as details of the four employed qubits can be found in the Appendix \ref{appendix:error_mitigation}.}\label{fig:penta_result}
\end{figure*}

The first experiment is carried out for the Hamiltonian parameters $x=0.8$ and $\tilde{m}=1.2$ [see Eq.~(\ref{Eq_Full_Encoded_Hamiltonian})].
This quark mass is large enough to organize the hadron spectrum in an intuitive way according to their mass: the vacuum, meson, tetraquark, and baryon-antibaryon states have energies near 0, $2\tilde m$, $4\tilde m$ and $6\tilde m$ respectively. For reference, Fig.~\ref{fig:results}a shows the spectrum and the composition of physical hadron states in terms of strong coupling states.
Since we choose the strong coupling baryon-antibaryon as initial state, quark-antiquark annihilation provides a direct connection to the tetraquark state, and indeed the collected data allows to observe this oscillation to dominate the time evolution as shown in Fig.~\ref{fig:results}b.
The experiment has been performed on the \texttt{ibm\_peekskill} device, where for each data point we run 280 (140, 140) repetitions in the case of $N_{\mathrm{T}}=4$ ($N_{\mathrm{T}}=2$, $N_{\mathrm{T}}=6$) for physics and mitigation runs respectively.
In Fig.~\ref{fig:results}c we perform a Bayesian analysis~\cite{kruschke2014} to extract the frequency of the oscillation and thus calculate the mass gap between the baryon-antibaryon state and the tetraquark state (see App.~\ref{app:bayes} for more details). We identify one frequency, where the point estimate is given by $\omega_{0} = (2\pi)\cdot0.262/[\tilde{m}t]$ with the highest density interval (HDI) of $(2\pi)\cdot[0.254,\,0.267]/[\tilde{m}t])$, which corresponds to the energy gap between $|\bar{B}B\rangle$ and $|\mathbb{T}\rangle$. The HDI is the interval where we find 95\% of the values during the sampling procedure. In Fig.~\ref{fig:results}c we plot 5000 samples from the posterior predictive distribution which within the 2.5 and 97.5 percent quantile agrees well with the collected data.

The second experiment is carried out on \texttt{ibm\_geneva} ($N_{\mathrm{T}}=4$, 560 repetitions) and \texttt{ibmq\_lima} ($N_{\mathrm{T}}$=8, 40 repetitions), employing a smaller quark mass $\tilde{m}=0.45$ but the same coupling constant $x=0.8$.
In this case
our quantum calculation of the real-time dynamics is able to reveal two dominant energy gaps, resulting in a beat frequency which is easily visible in Fig.~\ref{fig:results}e.  Applying the same Bayesian inference techniques, we extract two frequency components $\omega_1 = (2 \pi) \cdot 0.482 / [\tilde{m} t]$ with $\mathrm{HDI} = (2 \pi) \cdot [0.409,\, 0.545]/ [\tilde{m} t]$ and $\omega_2 = (2 \pi) \cdot 0.427/ [\tilde{m} t]$ where $\mathrm{HDI} = (2 \pi) \cdot [0.297,\, 0.502]/ [\tilde{m} t]$, which confirms the underlying physics.  In particular, the strong coupling $|\bar{B} B\rangle$ initial state is once again mixing with the strong coupling tetraquark, but this tetraquark is now a significant percentage of two physical eigenstates, as shown in Fig.~\ref{fig:results}d.

For our pentaquark study, we turn our attention to the subsector with baryon number $B=1$, and perform a Trotter time evolution under the four qubit Hamiltonian $\hat{H}^{(4)}$ derived in App.~\ref{app:trotter_penta}.
This scenario corresponds to a situation involving two heavy quarks and up to six light quarks, as shown in Fig.~\ref{fig:penta_result}a. We initialise the system in the state $|\mathbb{B}\rangle$, which corresponds to a baryon consisting of two heavy and one light quark [see Fig.~\ref{fig:penta_result}b]. The time evolution under $\hat{H}^{(4)}$ induces pair creation processes that cause the initial state $|\mathbb{B}\rangle$ to mix with the pentaquark $|\mathbb{P}\rangle$ [see Fig.~\ref{fig:penta_result}b-c] and a tetraquark-baryon pair $|\mathbb{T}b\rangle$, which consists of two heavy and four light particles. The latter is suppressed for our chosen parameter regime $x=3$, $\tilde{m}=0.1$, such that the observed dynamics is dominated by baryon-pentaquark oscillations. The corresponding experimentally calculated real time evolution of the particle number is shown in Fig.~\ref{fig:penta_result}d. This computation has been performed on \texttt{ibmq\_lima} and realizes $N_\mathrm{T}=2$ and $N_\mathrm{T}=4$ Trotter steps. Here, we performed around 160 repetitions to realize the full potential of the error mitigation technique.

\begin{figure}[t]
    \includegraphics[width=1\columnwidth]{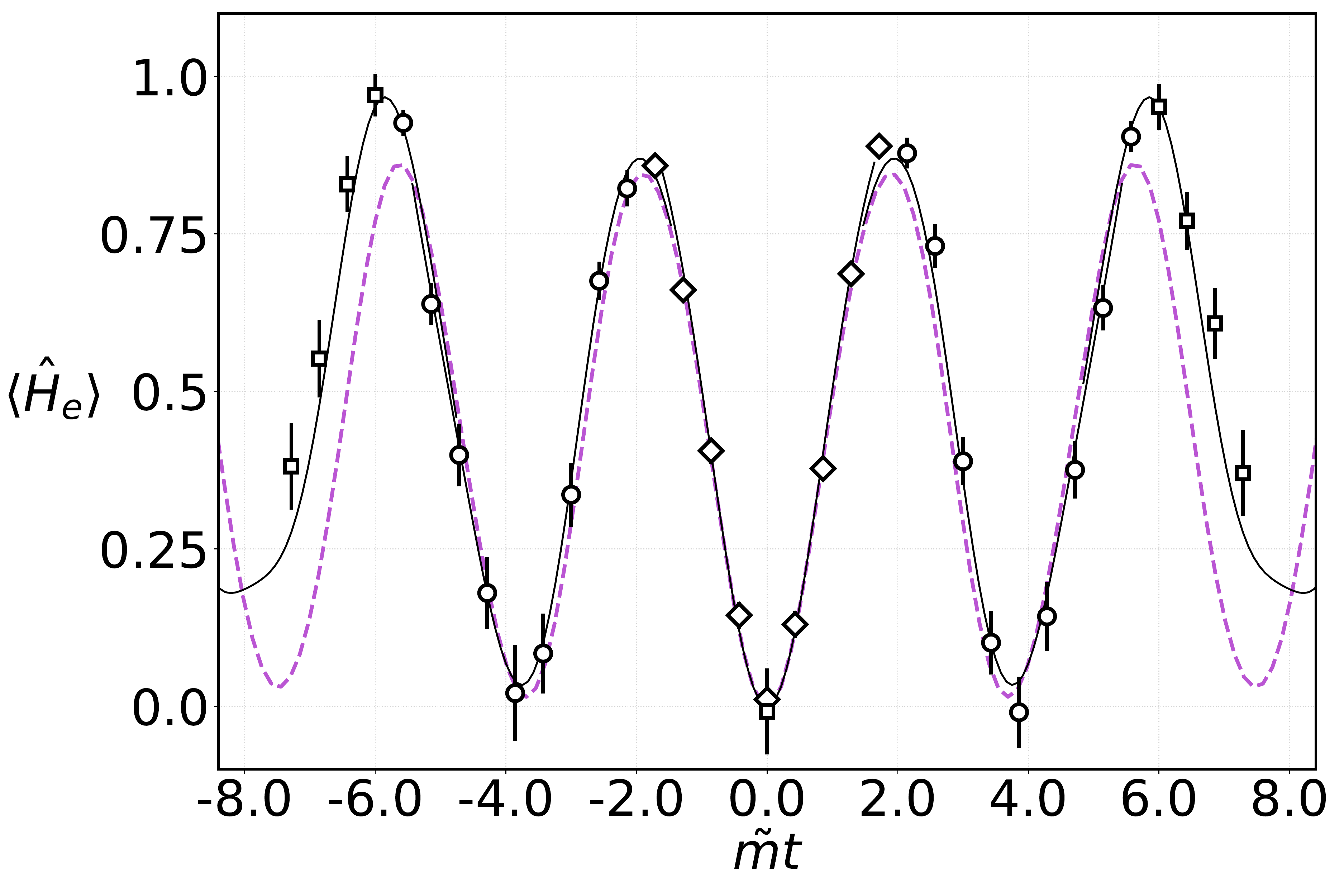} 
\caption{\textbf{Trotter evolution of the gauge fields} For the experiment shown in Fig.~\ref{fig:results}(b), where the parameters have been chosen as $\tilde{m}=1.2$, $x=0.8$, we reinterpret the measurement data and plot the energy contained in the gauge field of the time evolved state. 
The system is initialised in the zero flux baryonium state  $|\bar{B}B\rangle$ and subsequently evolved in time using the three qubits Hamiltonian derived in the Appendix. The dynamics is dominated by oscillations between the baryonium and tetraquark states containing one unit of electric flux. Full revival of the initial state is not observed due to contamination of the time evolved state with other available basis states. Legends for the data points and solid line are the same as in Fig.~\ref{fig:results}(b).} \label{fig:gauge_dynamic}
\end{figure}

\subsection{Time evolution of the gauge field}
\label{subsec:GaugeTimeEvo}
Our formulation of lattice QCD involves the explicit integration of Gauss law into the Hamiltonian, which enabled us to remove the gauge degrees of freedom from the description \cite{hamer_series_1997,kuhn_non-abelian_2015,martinez_real-time_2016,muschik_u1_2017}. 
However, we stress that this procedure does not restrict the access to the dynamics of the gauge field, which is now expressed in terms of the fermionic field operators. 
This can be deduced directly from Eq.~\eqref{electric_ham_qubit}, which expresses the energy of the gauge field solely via the non-Abelian charges. 
Note that this is indeed a direct consequence of the required gauge invariance of the system, realized by the interaction of the gauge field operators $\hat{U}_n$ and the fermionic field operators $\hat{\phi}_n$, in combination with Gauss law.
The latter implies direct constraints on the quantum states of the gauge field on all links surrounding each single site and the state of the matter at that site. 
In other words, in our case, the state of the gauge field can be deduced (up to a constant) as soon as the state of the matter is known.

The form of the gauge field Hamiltonian given in Eq.~\eqref{electric_ham_qubit} provides access to the gauge field on each link, when the corresponding charges $\hat{\mathbf{Q}}_n$ are measured.
As an explicit example, we show the time evolution of the total gauge field Hamiltonian $\langle\hat{H}_e^{(3)}\rangle$ in Fig.~\ref{fig:gauge_dynamic}.
The dynamics shown there is complementary to the evolution of the particle number shown in Fig.~\ref{fig:results}(b) and, importantly, it is deduced from the same data set. 
To be more precise, at time $\tilde{m}t=0$ the system is initialized in the  zero flux baryonium state $|\bar{B}B\rangle$. 
The subsequent time evolution is then dominated by oscillations between the baryonium and the tetraquark states, where the latter contains one unit of electric flux, which for SU(3) is equal to $4/3$.
At time $\tilde{m}t \approx 2$ we observe a maximum value for the electric energy.
Here, the state of the systems corresponds mostly to the tetraquark state (see Fig.~\ref{fig:results}(b)). 
Note that the value of the maximum is not exactly $4/3$ as mentioned above, since the time evolved state is a superposition still containing non negligible baryonium contributions.
Equivalently, the expectation value of the electric energy does not return exactly to zero at time $\tilde{m}t\approx 4$ as the time evolved state gets superimposed with the other contributions that are displayed in Fig.~\ref{fig:results}(a).  
\color{black}

\section{Discussion and Conclusion}

Recent observations of tetraquarks, pentaquarks, and other hadrons beyond the traditional mesons and baryons have sparked a great deal of theoretical activity \cite{Chen:2022asf}, with lattice gauge theory playing a central role in the determination of static properties.  To access time-dependent dynamics, we turn to the Hamiltonian approach on quantum computers.

Building on previous proposals~\cite{jordan_quantum_2012,jordan2014quantum,tong2021provably,ott2021scalable,kasper2017implementing,kasper2016schwinger,andrade2022engineering,davoudi2021toward,tagliacozzo_simulation_2013,mezzacapo_non-abelian_2015,bender_digital_2018,zohar_eliminating_2018,zache_dynamical_2019,kasper_non-abelian_2020,kasper_jaynescummings_2020,raychowdhury_loop_2020,davoudi_search_2021,davoudi2020towards,rico_so3_2018,laflamme2016cp,banerjee2022nematic,marcos2014two,kruckenhauser2022high,gonzalez-cuadra_hardware_2022,aidelsburger_cold_2022,dasgupta_cold-atom_2022,notarnicola2020real,shaw2020quantum,ciavarella2022some} and demonstrations in simpler gauge theories~\cite{klco_su2_2020,ARahman:2021ktn,atas2021,rahman2022,Illa:2022jqb,Fromm:2022vaj}, we have constructed a one-dimensional lattice gauge theory for QCD for two cases: one flavour of dynamical matter coupled to SU(3) gauge fields, and two flavours of matter, one of which corresponds to infinitely massive quarks.  Real-time oscillations of a tetraquark and a pentaquark with other hadron states are observed by running experiments on IBM Quantum hardware \cite{IBMQ}.  Specifically, we begin from a strong coupling eigenstate at time $t=0$ and see how the tetraquark and the pentaquark emerge. The success of these simulations required the use of recent advances in error mitigation \cite{rahman2022}.

Our approach is based on an elimination of the gauge degrees of freedom, with that physics being re-expressed as non-local interactions among matter fields. Future work will extend this methodology to two (and ultimately three) spatial dimensions using the methods developed in Ref.~\cite{haase_resource_2021}.

Another important route for generalisations is the extension of wider classes of simulated time evolutions to extract truly dynamical quantities. Interesting applications include time correlation functions and the ongoing quest to simulate particle collisions with quantum computers. 
Our simulation of SU(3) hadrons on a quantum computer accomplishes a key step on the path toward accessing increasingly relevant quantum computations for QCD.
\newline 


\section{Acknowledgements}
\noindent We are immensely grateful to Thomas Monz and Christian Sommer for sharing their important scientific insights and invaluable input to our project. We thank John Watrous for his support, and we are grateful for the IBM Quantum Researchers Program Access Award enabling the use of IBM Quantum services for this work. The views expressed are those of the authors, and do not reflect the official policy or position of IBM or the IBM Quantum team. 
This work has been supported by Transformative Quantum Technologies Program (CFREF), NSERC and the New Frontiers in Research Fund.
CM acknowledges the Alfred P. Sloan foundation for a Sloan Research Fellowship. JFH acknowledges the ERC Synergy Grant HyperQ (grant number 856432) and the BMBF project SPINNING (FKZ:13N16215).

\noindent


\appendix

\section{Theory}

\subsection{Gauge elimination and qubit formulation}
\label{app:qubit_formulation}
In the following, we discuss how to eliminate the gauge fields from the Hamiltonian and express it in terms of qubits only. 
Due to gauge invariance, the Hamiltonian in Eq.~(\ref{KSham}) commutes with the Gauss law operators (which generate the local gauge transformations) $\hat{G}_{n}^{a}\equiv \hat{L}_{n}^{a}-\hat{R}_{n-1}^{a}-\hat{Q}_{n}^{a}$, where $\hat{L}_{n}^{a}$ and $\hat{R}_{n-1}^{a}$ are the $a$-component (with $a=1,\dots,8$) of the left and right colour electric field defined on the link $n$ respectively.
For a non-Abelian gauge group, the right and left colour electric field are related via the adjoint representation $\hat{R}_{n}^{a}=(\hat{U}_n^\text{adj})_{ab}\hat{L}_{n}^{b}$, with 
$(\hat{U}_{n}^\text{adj})_{ab}=2\mathrm{Tr}\left[ \hat{U}_{n}T^{a}\hat{U}_{n}^{\dagger}T^{b}\right]$, where $T^{a}=\lambda^{a}/2$, and $\lambda^{a}\ (a=1,\dots,8)$ are generators of the SU(3) Lie algebra and are given by the Gell-Mann matrices \cite{griffiths2020introduction}.
The Hamiltonian also commutes with the redness operator $\hat{\mathcal{R}} =\sum_{n=1}^{N} \hat{\phi}_{n}^{1 \dagger}\hat{\phi}_{n}^{1}-N/2$, the greenness operator $\hat{\mathcal{G}} =\sum_{n=1}^{N} \hat{\phi}_{n}^{2 \dagger}\hat{\phi}_{n}^{2}-N/2$, and the blueness operator $\hat{\mathcal{B}} =\sum_{n=1}^{N} \hat{\phi}_{n}^{3 \dagger}\hat{\phi}_{n}^{3}-N/2$ which measure the matter-antimatter imbalance of a specific colour. It is however more convenient to combine these three operators into a single gauge-invariant one which measures the matter-antimatter imbalance irrespective of the colour. We therefore define the baryon number operator as
\begin{equation}
 \hat{B}=\frac{1}{3}\left(\hat{\mathcal{R}}+\hat{\mathcal{G}}+\hat{\mathcal{B}}\right). \label{methods_defbaryonnum}
\end{equation}

To simulate time dynamics 
on a quantum computer, we transform the fermionic Hamiltonian in Eq.~(\ref{KSham}) into one involving only qubit degrees of freedom. The transformation is achieved in two steps. We start by first eliminating the gauge fields following  \cite{kuhn_non-abelian_2015,atas2021}. The resulting dimensionless Hamiltonian reads
\begin{align}
\hat{H} = \notag & \frac{1}{2} \sum_{n=1}^{N-1} \sum_{i=1}^{3}\left( \hat{\phi}_{n}^{i \dagger}\hat{\phi}_{n+1}^{i} + \operatorname{H.c.}\right) \\ 
&+ \tilde{m} \sum_{n=1}^{N}\sum_{i=1}^{3} (-1)^{n} \hat{\phi}_{n}^{i \dagger} \hat{\phi}_{n}^{i} + \frac{1}{2x} \sum_{n=1}^{N-1}\left( \sum_{m\leq n}\boldsymbol{\hat{{Q}}_{m}}\right)^{2} \label{appendix_KSham_gauge_free},
\end{align}
where $\tilde{m}=am$  and $x=1/g^{2}a^2$. The last term represents the colour electric energy of the system and is expressed in terms of the non-Abelian charges at site $n$,
\begin{equation}
  \hat{Q}_{n}^{a}=\sum_{i,j=1}^{3}\hat{\phi}_{n}^{i\dagger}(T^{a})_{ij}\hat{\phi}_{n}^{j}. \label{methods_non_abelian_charges}
\end{equation}
In a second step, we triple the size of the lattice to define $3N$ new sites and distribute the colour components of the fermionic field among them by defining the single component fields $\hat{\phi}_{n}^{i}=\hat{\psi}_{3n-3+i}$ with $n=1,2,\dots,N$ and $i=1,2,3$ [see Fig.~\ref{fig:tetraquark}a].

We then perform a generalised Jordan-Wigner transformation on the single component fermionic fields $\hat{\psi}_{n}$ \cite{JordanWigner}
\begin{equation}
    \hat{\psi}_{n}=\left(\prod_{l<n} s_{l} \hat{\sigma}_{l}^{z}\right)\hat{\sigma}_{n}^{-}, \quad \hat{\psi}_{n}^{\dagger}=\left(\prod_{l<n} s_{l} \hat{\sigma}_{l}^{z}\right)\hat{\sigma}_{n}^{+},
\end{equation}
where $s_{l}$ are phase factors that we choose equal to $+1$  on antimatter sites and  $-1$ on matter sites.
This choice is convenient because it matches the standard colour notation, such as $(\bar rr+\bar gg+\bar bb)/\sqrt{3}$ for a meson.
After the Jordan-Wigner transformation, the kinetic Hamiltonian in terms of qubits is given by Eq.~(\ref{kinetic_ham_qubit})
while the mass Hamiltonian is given by Eq.~(\ref{mass_ham_qubit}). 
In the qubit formulation, the non-Abelian charges defined in Eq.~(\ref{methods_non_abelian_charges}) read 
\begin{align}
    \hat{Q}_{n}^{1}&=\frac{(-1)^{n}}{2}\left( \hat{\sigma}_{3n-2}^{+}\hat{\sigma}_{3n-1}^{-}+\operatorname{H.c.}\right)\label{qubit_nonabeliancharge1}, \\
    \hat{Q}_{n}^{2}&=\frac{i(-1)^{n}}{2}\left( \hat{\sigma}_{3n-1}^{+}\hat{\sigma}_{3n-2}^{-}-\operatorname{H.c.}\right)\label{qubit_nonabeliancharge2},\\
    \hat{Q}_{n}^{3}&=\frac{1}{4}\left( \hat{\sigma}_{3n-2}^{z}-\hat{\sigma}_{3n-1}^{z}\right)\label{qubit_nonabeliancharge3},\\
    \hat{Q}_{n}^{4}&=-\frac{1}{2}\left( \hat{\sigma}_{3n-2}^{+}\hat{\sigma}_{3n-1}^{z}\hat{\sigma}_{3n}^{-}+\operatorname{H.c.}\right)\label{qubit_nonabeliancharge4}, \\
    \hat{Q}_{n}^{5}&=\frac{i}{2}\left( \hat{\sigma}_{3n-2}^{+}\hat{\sigma}_{3n-1}^{z}\hat{\sigma}_{3n}^{-}-\operatorname{H.c.}\right)\label{qubit_nonabeliancharge5}, \\
    \hat{Q}_{n}^{6}&=\frac{(-1)^n}{2}\left( \hat{\sigma}_{3n-1}^{+}\hat{\sigma}_{3n}^{-}+\operatorname{H.c.}\right)\label{qubit_nonabeliancharge6}, \\
    \hat{Q}_{n}^{7}&=\frac{i(-1)^n}{2}\left( \hat{\sigma}_{3n}^{+}\hat{\sigma}_{3n-1}^{-}-\operatorname{H.c.}\right)\label{qubit_nonabeliancharge7},\\
    \hat{Q}_{n}^{8}&=\frac{1}{4\sqrt{3}}\left( \hat{\sigma}_{3n-2}^{z}+\hat{\sigma}_{3n-1}^{z}-2\hat{\sigma}_{3n}^{z}\right)\label{qubit_nonabeliancharge8}.
\end{align}
The colour electric field Hamiltonian can be obtained by inserting the qubit expressions for the non-Abelian charges in Eqs.~(\ref{qubit_nonabeliancharge1}-\ref{qubit_nonabeliancharge8}) into the colour electric term in Eq.~(\ref{appendix_KSham_gauge_free}). We obtain
\begin{align}
    \hat{H}_{e}=&\notag \frac{1}{3}\sum_{n=1}^{N-1}(N-n)\times \\ 
    \notag & \left( 3-\hat{\sigma}_{3n-2}^{z}\hat{\sigma}_{3n-1}^{z}-\hat{\sigma}_{3n-2}^{z}\hat{\sigma}_{3n}^{z}-\hat{\sigma}_{3n-1}^{z}\hat{\sigma}_{3n}^{z}\right)\\
   \notag &+\sum_{n=1}^{N-2}\sum_{m=n+1}^{N-1}\left[ (N-m)\left( \hat{\sigma}_{3n-2}^{+}\hat{\sigma}_{3n-1}^{-}\hat{\sigma}_{3m-1}^{+}\hat{\sigma}_{3m-2}^{-}\right. \right. \\
   \notag & +\left. \hat{\sigma}_{3n-1}^{+}\hat{\sigma}_{3n}^{-}\hat{\sigma}_{3m-1}^{-}\hat{\sigma}_{3m}^{+}+\operatorname{H.c.}\right)(-1)^{n+m} \\
   \notag & +(N-m)\left(\hat{\sigma}_{3n-2}^{+}\hat{\sigma}_{3n-1}^{z}\hat{\sigma}_{3n}^{-}\hat{\sigma}_{3m-2}^{-}\hat{\sigma}_{3m-1}^{z}\hat{\sigma}_{3m}^{+} +\operatorname{H.c.}\right) \\
   \notag &-\frac{1}{12}(N-m) \hat{\sigma}_{3m-2}^{z}(\hat{\sigma}_{3n-1}^{z}+\hat{\sigma}_{3n}^{z}-2\hat{\sigma}_{3n-2}^{z}) \\
   \notag &-\frac{1}{12}(N-m) \hat{\sigma}_{3m-1}^{z}(\hat{\sigma}_{3n}^{z}+\hat{\sigma}_{3n-2}^{z}-2\hat{\sigma}_{3n-1}^{z})\\
    & \left. -\frac{1}{12}(N-m) \hat{\sigma}_{3m}^{z}(\hat{\sigma}_{3n-2}^{z}+\hat{\sigma}_{3n-1}^{z}-2\hat{\sigma}_{3n}^{z}) \right], \label{qubit_electric_general}
\end{align}
which exhibits long range spin-spin interaction as a direct consequence of the gauge elimination. The baryon number operator is proportional to the total magnetization of the system in the qubit encoding 
\begin{equation}
    \hat{B}=\frac{1}{6}\sum_{n=1}^{3N}\hat{\sigma}_{n}^{z}. \label{qubit_baryon_number}
\end{equation}

\subsection{Hamiltonian for $N=2$ and reduction to three qubits}
\label{app:reduction_to_three_qubits}
We are interested in a basic building block consisting of $N=2$ lattice sites. The model is then described by a chain with six qubits. The terms in the Hamiltonian read
\begin{align}
    \hat{H}_{kin}&=-\frac{1}{2}\left( \hat{\sigma}_{1}^{+}\hat{\sigma}_{2}^{z}\hat{\sigma}_{3}^{z} \hat{\sigma}_{4}^{-}-\hat{\sigma}_{2}^{+}\hat{\sigma}_{3}^{z}\hat{\sigma}_{4}^{z} \hat{\sigma}_{5}^{-}+\hat{\sigma}_{3}^{+}\hat{\sigma}_{4}^{z}\hat{\sigma}_{5}^{z} \hat{\sigma}_{6}^{-}+\operatorname{H.c.}\right), \label{6qubit_kinetic}\\
    \hat{H}_{m}&=\frac{1}{2}\left(6- \hat{\sigma}_{1}^{z}-\hat{\sigma}_{2}^{z}-\hat{\sigma}_{3}^{z}+\hat{\sigma}_{4}^{z}+\hat{\sigma}_{5}^{z}+\hat{\sigma}_{6}^{z} \right), \label{6qubit_mass}\\
    \hat{H}_{e}&=\frac{1}{3}\left(3-\hat{\sigma}_{1}^{z}\hat{\sigma}_{2}^{z}-\hat{\sigma}_{1}^{z}\hat{\sigma}_{3}^{z} -\hat{\sigma}_{2}^{z}\hat{\sigma}_{3}^{z}\right)\label{6qubit_elec}.
\end{align}
In the sector with baryon number $B=0$ (\textit{i.e.} with zero matter-antimatter imbalance), the three terms composing the Hamiltonian commute with the following operator 
\begin{equation}
    \hat{CP}=\prod_{n=1}^{3}\hat{\sigma}_{n}^{x}\hat{\sigma}_{7-n}^{x}\hat{W}_{n,7-n},
\end{equation}
where $\hat{W}_{n,n'}$ is the SWAP unitary operator between qubits $n$ and $n'$. This symmetry corresponds to the composition of a spatial reflection ($\hat{P}$) with respect to the middle of the chain followed by a charge conjugation operation ($\hat{C}$) which flips the spins. Local spin operators $\hat{\sigma}_{n}^{a}$ transform as $(\hat{CP})^{\dagger}\hat{\sigma}_{n}^{a}\hat{CP}=(\hat{\sigma}^{x}\hat{\sigma}^{a}\hat{\sigma}^{x})_{7-n}$ under the $\hat{CP}$ operation with $a=x,y,z$ and $n=1,2,\dots,6$. It is thus clear that a convenient  basis is  the one spanned by states of the form  $|\Psi\rangle = \sum_{ijk} c_{ijk} |i\rangle_1|j\rangle_2 |k\rangle_3 \otimes \hat{\sigma}_{4}^x \hat{\sigma}_{5}^x\hat{\sigma}_{6}^x|i\rangle_4|j\rangle_5|k\rangle_6$, which are invariant under the $\hat{CP}$ operation and are $B=0$ eigenstates. Working with this basis, the state of the last three qubits is determined by the state of the first three.
As a direct consequence, we can encode the states by using only the first three qubits  (\textit{i.e.} the state of the antimatter) rather than six. The reduced three-qubit Hamiltonian reads
\begin{align}\label{HkinXZ}
    \hat{H}_{kin}^{(3)}&=-\frac{1}{2}\left( \hat{\sigma}_{1}^{x}\hat{\sigma}_{2}^{z}\hat{\sigma}_{3}^{z}+\hat{\sigma}_{1}^{z}\hat{\sigma}_{2}^{x}\hat{\sigma}_{3}^{z}+\hat{\sigma}_{1}^{z}\hat{\sigma}_{2}^{z}\hat{\sigma}_{3}^{x}\right),\\
    \hat{H}_{m}^{(3)}&=3-\hat{\sigma}_{1}^{z}-\hat{\sigma}_{2}^{z}-\hat{\sigma}_{3}^{z},\\
    \hat{H}_{e}^{(3)}&=\frac{1}{3}\left(3-\hat{\sigma}_{1}^{z}\hat{\sigma}_{2}^{z}-\hat{\sigma}_{1}^{z}\hat{\sigma}_{3}^{z} -\hat{\sigma}_{2}^{z}\hat{\sigma}_{3}^{z}\right), \label{HeXZ}
\end{align}
and the time evolution is obtained using the Hamiltonian
\begin{equation}
    \label{eq:three_qubit_ham}
    \hat{H}^{(3)}=\hat{H}_{kin}^{(3)}+\tilde{m}\hat{H}_{m}^{(3)}+\frac{1}{2x}\hat{H}_{e}^{(3)}.
\end{equation}

\subsection{Inclusion of static charges}
\label{app:inclusion_of_static}
Static charges can be effectively incorporated in our model using two flavours of quarks. We use light quarks with mass $\tilde{m}$ for the dynamical charges and heavy quarks with mass $\tilde{M}\gg \tilde{m}$ to represent the external static charges. By definition, the static charges do not enter into the kinetic term, and for simplicity, we remove their contribution from the mass Hamiltonian by a shift in the zero energy definition. The Gauss law must be modified in order to take the colour electric field created by the static charges into account: the non-Abelian charges appearing in the colour electric Hamiltonian in Eq.~(\ref{electric_ham_qubit}) are replaced by $\hat{\bm{\tilde{Q}}}_{n}=\hat{\mathbf{Q}}_{n}^{L}+\hat{\mathbf{Q}}_{n}^{H}$ where $\hat{\mathbf{Q}}_{n}^{L}$ are the light dynamical charges defined in Eq.~(\ref{methods_non_abelian_charges}) and $\hat{\mathbf{Q}}_{n}^{H}$ is the heavy static source distribution at cell $n$. We choose the following configuration for the external charges: we place one static anticharge at cell $n_{1}$ (odd cell) and a static charge at cell $n_{2}$ (even cell). The static charge distribution is thus given by $\hat{\mathbf{Q}}_{n}^{H}=\hat{\mathbf{q}}_{n_{1}}\delta_{n,n_{1}}+\hat{\mathbf{q}}_{n_{2}}\delta_{n,n_{2}}$ where $\hat{q}_{n}^{a}=\sum_{i,j=1}^{3}\hat{\eta}_{n_{}}^{i\dagger}(T^{a})_{ij}\hat{\eta}_{n}^{j}$ and $\hat{\eta}_{n}$ is a three colour component fermion fields associated with the static charges. Expanding the square in the colour electric field Hamiltonian $\hat{H}_{e}^{(q)}=\sum_{n=1}^{N-1}\left( \sum_{m\leq n} \hat{\bm{\tilde{Q}}}_{m}\right)^2$, we find that the colour electric field Hamiltonian in presence of external charges $\hat{H}_{e}^{(q)}$ is the sum of three contributions
\begin{equation}
    \hat{H}_{e}^{(q)}=\hat{H}_{e}^{(q=0)}+\hat{H}_{e}^{(qq)}+\hat{H}_{e}^{(qQ)}
\end{equation}
where $\hat{H}_{e}^{(q=0)}$ is the colour electric field generated by the light dynamical charges alone and is given by Eq.~(\ref{electric_ham_qubit}), the second term
\begin{align}
    \notag \hat{H}_{e}^{(qq)}=\sum_{a=1}^{8}&\left[ (\hat{q}_{n_{1}}^{a})^2(N-n_{1})+(\hat{q}_{n_{2}}^{a})^2(N-n_{2}) \right. \\
& \left. +2\hat{q}_{n_{1}}^{a}\hat{q}_{n_{2}}^{a} (N-n_{2})  \right] \label{external_charges_elec_int},
\end{align}
describes the interaction between the two external charges (we assume $n_{1}<n_{2}$ without loss of generality), and 
\begin{align}
    \notag \hat{H}_{e}^{(qQ)}=2\sum_{a=1}^{8}\sum_{n=1}^{N-1}\hat{Q}_{n}^{a}&\left(\hat{q}_{n_{1}}^{a}(N-\max(n_{1},n)) \right. \\
    &\left. +\hat{q}_{n_{2}}^{a}(N-\max(n_{2},n))\right) \label{external_dynamical_elec_int}
\end{align}
is the interaction between the external and the dynamical charges. In order to describe the system in terms of qubits, we follow the same steps as before and use $3N$ qubits to describe the dynamical charges and add six extra qubits at the end of the chain to describe the external charges degrees of freedom (see Fig.~\ref{fig:penta_result}a). For the expression of the external charges in terms of qubits, we can use the definition Eqs.~(\ref{qubit_nonabeliancharge1}-\ref{qubit_nonabeliancharge8}) with $n=N+1$ for $\hat{q}_{n_{1}}^{a}$ and $n=N+2$ for $\hat{q}_{n_{2}}^{a}$.
In total, we thus need $3N+6$ qubits to describe the system with two external charges. 

\subsection{Trotter Hamiltonian to study pentaquarks} 
\label{app:trotter_penta}
To study the properties of pentaquark states, consisting of four quarks and one antiquark, we consider the minimal system size which can host such a state. We therefore work with $N=2$ for the light dynamical charges, one static anticharge at site $n_{1}=1$, and one static charge at site $n_{2}=2$.  
We arrange the light and heavy quarks in the unit cell containing 12 qubits as shown in Fig.~\ref{fig:penta_result}a, with the first six qubits describing the light dynamical quarks, and the last six qubits encoding the state of the heavy quarks. We define the total baryon number $\hat{B}=\hat{B}_{L}+\hat{B}_{H}$ of the system as the sum of the light quarks baryon number $\hat{B}_{L}$ and the heavy quark baryon number $\hat{B}_{H}$, 
\begin{align}
    \hat{B}_{L}&=\frac{1}{6}\sum_{n=1}^{3N}\hat{\sigma}_{n}^{z},\\
        \hat{B}_{H}&=\frac{1}{6}\sum_{n=3N+1}^{3N+6}\hat{\sigma}_{n}^{z}.
\end{align}

Since we are interested in a pentaquark state, we restrict ourselves to the sector with total baryon number $B=1$. 
We choose to study a pentaquark state that contains a heavy diquark. This choice of pentaquark translates into two constraints. First, the absence of anticharge colours on the heavy antiquark sites encoded in qubits 7, 8 and 9 translates into fixing these spins in the vacuum state for this part of the lattice $\ket{vac}=\ket{\uparrow_{7}\uparrow_{8}\uparrow_{9}}$. 
Second, imposing that the heavy quark matter sites (encoded in qubits 10, 11 and 12) represent a diquark state, requires the last three qubits 10, 11 and 12 to be in a state with total cell magnetisation $M=1$. 



We perform a time evolution experiment starting from the baryon state (consiting of one light quark and one heavy diquark) which reads in spin formulation 
\begin{align}
    \notag \ket{\Psi_{B}}=\frac{1}{\sqrt{3}}&\left( \ket{\uparrow\uparrow\uparrow \uparrow\downarrow\downarrow}_{L}\ket{\uparrow\uparrow\uparrow \downarrow \uparrow \uparrow}_{H}-\ket{\uparrow\uparrow\uparrow \downarrow\uparrow\downarrow}_{L}\ket{\uparrow\uparrow\uparrow \uparrow \downarrow \uparrow}_{H} \right. \\ 
    & -\left. \ket{\uparrow\uparrow\uparrow \downarrow\downarrow\uparrow}_{L}\ket{\uparrow\uparrow\uparrow \uparrow \uparrow \downarrow}_{H} \right),\label{baryon_initial_state_pentaexp}
\end{align}
where the first ket with subscript $L$ represents the light quarks while the second one with subscript $H$ encodes the heavy quarks. 
Note that the state of the first three qubits in the heavy quark ket is fixed (due to the absence of heavy anticharges) and the state of the last three qubits in the heavy quark ket correponds to the flipped version of the last three qubits in the light quark ket. As a consequence, the heavy quark sites do not need to be included in the quantum simulation. This is also reflected in the Hamiltonian governing the dynamics, which is given in Eqs.~(\ref{6qubit_kinetic}-\ref{6qubit_elec}) and involves only the first six qubits. 

Since the state of the light quarks is sufficient to reconstruct the full state of the system, we rewrite the baryon initial state in terms of the light quark ket only
\begin{equation}
     \ket{\Psi_{B}}=\frac{1}{\sqrt{3}}\left( \ket{\uparrow\uparrow\uparrow \uparrow\downarrow\downarrow}_{L}-\ket{\uparrow\uparrow\uparrow \downarrow\uparrow\downarrow}_{L} - \ket{\uparrow\uparrow\uparrow \downarrow\downarrow\uparrow}_{L} \right). \label{baryon_initial_light_quark}
\end{equation}
Each of the three terms in the superposition corresponds to a different colour sector. The first ket has a light redness number $\hat{\cal{R}}_{L}=(\hat{\sigma}_{1}^{z}+\hat{\sigma}_{4}^{z})/2$ equal to 1. Similarly, the second ket in the expansion has a light greeness number $\hat{\cal{G}}_{L}=(\hat{\sigma}_{2}^{z}+\hat{\sigma}_{5}^{z})/2$ equal to 1, while the last term has a light blueness number defined as $\hat{\cal{B}}_{L}=(\hat{\sigma}_{3}^{z}+\hat{\sigma}_{6}^{z})/2$ equal to 1. Since the Hamiltonian in Eqs.~(\ref{6qubit_kinetic}-\ref{6qubit_elec}) preserves the light redness, greenness and blueness quantum numbers, the time evolution will not mix the different colour sectors. As a consequence, it is sufficient to time evolve only one of the kets in Eq.~(\ref{baryon_initial_light_quark}) rather than applying the time evolution operator to the superposition.
This observation allows us to significantly decrease the complexity of our Trotter protocol. 
We thus use $\ket{\Psi_{B}^{(r)}}=\ket{\uparrow\uparrow\uparrow \uparrow\downarrow\downarrow}_{L}$ as our initial state, where the superscript indicates that this state belongs to the red sector. Furthermore, since qubit 1 and qubit 4 are not affected by the time evolution, the dynamics is effectively governed by the four-qubit Hamiltonian $\hat{H}^{(4)}=\hat{H}_{kin}^{(4)}+(1/2x)\hat{H}_{e}^{(4)}+\tilde{m}\hat{H}_{m}^{(4)}$ with
\begin{align}
    \hat{H}_{kin}^{(4)}&=-\frac{1}{2}\left(-\hat{\sigma}_{2}^{+}\hat{\sigma}_{3}^{z} \hat{\sigma}_{5}^{-}+\hat{\sigma}_{3}^{+}\hat{\sigma}_{5}^{z} \hat{\sigma}_{6}^{-}+\operatorname{H.c.}\right), \label{4qubit_kinetic}\\
    \hat{H}_{m}^{(4)}&=\frac{1}{2}\left(6 -\hat{\sigma}_{2}^{z}-\hat{\sigma}_{3}^{z}+\hat{\sigma}_{5}^{z}+\hat{\sigma}_{6}^{z} \right), \label{4qubit_mass}\\
    \hat{H}_{e}^{(4)}&=\frac{1}{3}\left(3-\hat{\sigma}_{2}^{z}-\hat{\sigma}_{3}^{z} -\hat{\sigma}_{2}^{z}\hat{\sigma}_{3}^{z}\right)\label{4qubit_elec}.
\end{align} 
We find that a Trotter step time evolution of the four qubit Hamiltonian can be realised using 18 $CNOT$ gates. 
In our time evolution experiments (see Fig.~\ref{fig:penta_result}), we measure the total particle number of the time evolved state $\ket{\Psi(t)}=e^{-it\hat{H}}\ket{\Psi_{B}}$,
\begin{align}
  \notag  \langle \hat{\mathcal{N}}(t)\rangle&=\bra{\Psi(t)}\hat{H}_{m}\ket{\Psi(t)}+2\\
    &=\bra{\Psi^{(r)}(t)}\hat{H}_{m}^{(4)}\ket{\Psi^{(r)}(t)}+3,
\end{align}
where $\ket{\Psi^{(r)}(t)}=e^{-it\hat{H}^{(4)}}\ket{\Psi_{B}^{(r)}}$ is the time evolved state within the red sector.

\section{Experimental details}
\subsection{Trotter evolution}
\label{app:trotter}
\begin{figure}[ht]
\includegraphics[width=0.48\textwidth]{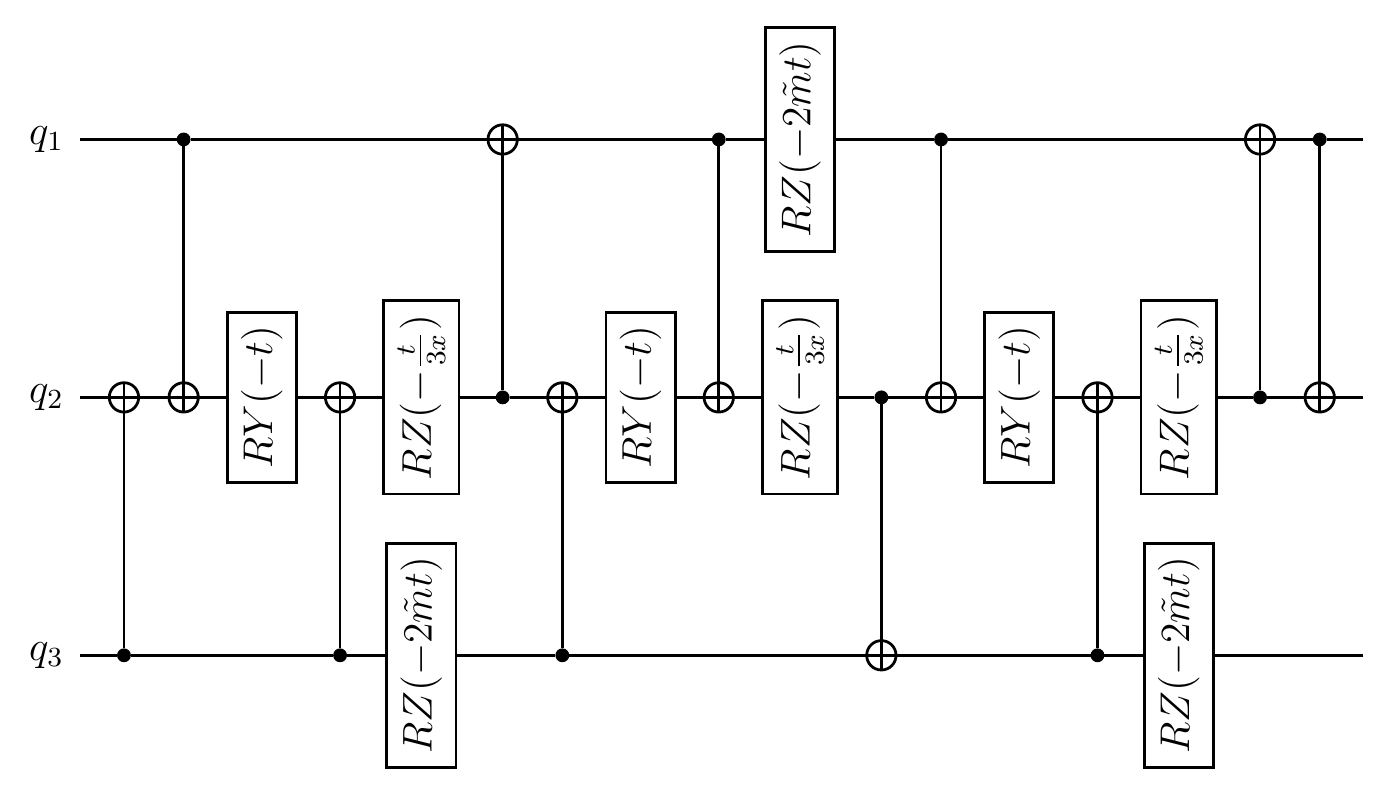} 
\caption{The first half of a Trotter step is shown in this figure.
The second half is identical except for the interchange of two qubits: $q_1 \leftrightarrow q_3$.
Note that a pair of $CNOT$ gates can be canceled where the two halves meet.}
\label{fig:trotter_circuit}
\end{figure}

In this section, we provide the circuit implementing one Trotter step for the tetraquark experiments. 
Although the Hamiltonian of Eqs.~(\ref{HkinXZ}-\ref{HeXZ}) is expressed in terms of Pauli $X$ and $Z$ gates, a simple rotation to $Y$ and $Z$ gates allows for more cancellations among $CNOT$ gates.  This is especially valuable on hardware that does not provide all-to-all connectivity among the qubits.  The first half of our first-order Trotter step is displayed in Fig.~\ref{fig:trotter_circuit} and, to match the available hardware, it does not use any entangling gates directly between $q_1$ and $q_3$.  The second half of the Trotter step is the same except for a relabeling of $q_1 \leftrightarrow q_3$.

Let us provide details about the procedure to obtain the Trotter circuit in Fig.~\ref{fig:trotter_circuit}.  We start from the three qubit Hamiltonian in Eqs.~(\ref{HkinXZ})-(\ref{HeXZ}) and perform a rotation to transform the $X$ Pauli matrices into $Y$ Pauli matrices. The unitary operator realising this rotation is given by
\begin{equation}
\hat{U}=\exp\left(i\frac{\pi}{2}(Z_{1}+Z_{2}+Z_{3})\right),
\end{equation}
and transforms the three qubit Hamiltonian into 
\begin{align}
    \hat{H}_{kin}&=-\frac{1}{2}\left(Y_{1}Z_{2}Z_{3}+Z_{1}Y_{2}Z_{3}+Z_{1}Z_{2}Y_{3} \right),\\
    \hat{H}_{m}&=3-Z_{1}-Z_{2}-Z_{3},\\
    \hat{H}_{e}&=\frac{1}{3}\left(3-Z_{1}Z_{2}-Z_{1}Z_{3}-Z_{2}Z_{3}\right).
\end{align}
Note that the mass term and electric term are not affected by the rotation as they only depend on $Z$ Pauli operators. Taking the time evolution operator of each term in the Hamiltonian, we see that a first order Trotterization produces terms of the form 
\begin{equation}
    \exp \left(i\frac{t}{2} Z_{1}Y_{2}Z_{3}\right), \quad \exp\left(i\frac{t}{6x} Z_{1}Z_{2}\right),
\end{equation}
which can be implemented using the following identities
\begin{equation}
    \exp \left(i\theta Z_{1}Y_{2}Z_{3}\right)= CX_{12}CX_{32}RY_{2}(-2\theta)CX_{32}CX_{12},
\end{equation}
and 
\begin{equation}
    \exp \left(i\theta Z_{1}Z_{2}\right)= CX_{12}RZ_{2}\left(-2\theta\right)CX_{12},
\end{equation}
where $CX_{ij}$ is a CNOT gate with control on qubit $i$ and target on qubit $j$ and, $RZ_{j}(\theta)=e^{-i\theta/2 Z_{j}}$ and  $RY_{j}(\theta)=e^{-i\theta/2 Y_{j}}$ are usual single qubit rotation gates. By defining
\begin{equation}
    U_{ijk}=e^{it/2Z_{i}Y_{j}Z_{k}}e^{it\tilde{m}Z_{k}}e^{it/6xZ_{i}Z_{j}}
\end{equation}
and using the identities above, the circuit can be obtained by implementing the following unitary operation
\begin{equation}
    U_{Trotter}=U_{123}U_{312}U_{231}.
\end{equation}

The error of our first order Trotter scheme $\hat{U}_T(t)$ of the true evolution induced by $\hat{H}=\sum_j\hat{h}_j$ can be estimated employing the relation (with spectral norm)\cite{Childs2021Theory}
\begin{equation}
    ||\hat{U}_T(t) - \mathrm{e}^{-it\hat{H}}|| = \mathcal{O}(\alpha t^2).
\end{equation}
Here, the $\hat{h}_j$ are specified by the components of the qubit Hamiltonians and
\begin{equation}
    \alpha = \sum_{j,k} ||[\hat{h}_j,\hat{h}_k]||.
\end{equation}
When applying $n$ Trotter steps, the total error of that evolution is then given as
\begin{equation}
    ||\hat{U}^n_T(t/n) - \mathrm{e}^{-it\hat{H}}|| = \mathcal{O}\left(\frac{\alpha^2t^2}{n}\right).
\end{equation}
For the cases examined in this work, we find that
\begin{eqnarray}
    \alpha_{\mathrm{tetra}} &=& 8|\tilde{m}| + \frac{8}{3|x|} \\
    \alpha_{\mathrm{penta}} &=& 2 + \alpha_{\mathrm{tetra}}.
\end{eqnarray}

\subsection{Scalability and resource estimation}
\label{app:scalability}
The experiments carried out were done for the minimal number of sites $N=2$, to be implemented on current quantum computers. Here we estimate the number of $CNOT$ gates needed to implement time evolution for larger system sizes $N$. 
For simplicity, we assume here all to all connectivity for the quantum hardware, as available in trapped ion quantum computers, and estimate the number of $CNOT$ operations reuqired to simulate one Trotter step for $N$ lattice sites (corresponding to $3N$ qubits). 
Consider a Pauli string $\hat{P}=\hat{P}_{1}\hat{P}_{2}\dots \hat{P}_{m}$ of length $m$ with $\hat{P}_{i}=\lbrace X,Y,Z\rbrace$. The unitary $e^{-it\hat{P}}$ can be implemented using $2(m-1)$ $CNOT$ gates. 
This allows us to estimate the number of $CNOT$ operations by counting the number of Pauli strings and their length in each term of the Hamiltonian in Eq.~(\ref{Eq_Full_Encoded_Hamiltonian}).  
The kinetic term in Eq.~(\ref{kinetic_ham_qubit}) has $3(N-1)$ terms of the form $\hat{\sigma}^{+}\hat{\sigma}^{z}\hat{\sigma}^{z}\hat{\sigma}^{-}+H.c.$ which corresponds to two Pauli strings $XZZX$ and $YZZY$ of length $4$. Thus, $36(N-1)$ $CNOT$ gates are needed to implement one Trotter step for the kinetic term. 
The mass Hamiltonian in Eq.~(\ref{mass_ham_qubit}) can be implemented using only $Z$ rotations and does not necessitate the use of entangling gates. The colour electric Hamiltonian in Eq.~(\ref{qubit_electric_general})
is the most costly in terms of entangling gates as it involves a number of terms growing quadratically with the size of the system $N$. The first term in Eq.~(\ref{qubit_electric_general}) has $3(N-1)$ Pauli strings of the form $ZZ$ which amount for $6(N-1)$ $CNOT$ gates. The double sum also involves $ZZ$ terms, of which there are $9(N^2-3N+2)/2$, therefore contributing with $9(N^2-3N+2)$ $CNOT$ gates. 
The $4$-body term $\hat{\sigma}^{+}\hat{\sigma}^{-}\hat{\sigma}^{+}\hat{\sigma}^{-}$ generates $9$ Pauli strings of length $4$ ($XXXX,XYXX,XXYY,\dots$), thus we need $54(N^2-3N+2)$ $CNOT$ gates to simulate such a term. Finally, the $6$-body term $\hat{\sigma}^{+}\hat{\sigma}^{z}\hat{\sigma}^{-}\hat{\sigma}^{-}\hat{\sigma}^{z}\hat{\sigma}^{+}$ can be decomposed into 8 Pauli strings of length $6$ and can be implemented using $40(N^{2}-3N+2)$ $CNOT$ gates. 
In total, we thus find that the number of $CNOT$ gates needed to implement one Trotter step of time evolution under the qubit Hamiltonian in Eq.~(\ref{Eq_Full_Encoded_Hamiltonian}) is given by $103N^2-267N+164$, and grows at most quadratically in the number of lattice sites $N$. 

\begin{table*}[]
    \centering
    \begin{tabular}{l|c|cccc|cccc|c|c}
         System name & qubits & \multicolumn{4}{c|}{$T_1\,(\mathrm{\mu s})$} & \multicolumn{4}{c|}{$T_2\,(\mathrm{\mu s})$} & largest Pauli error & largest $CNOT$ error\\ \hline \hline
        \texttt{ibmq\_lima} & 0, 1, 2 & 106.58 & 59.93 & 153.32 & &  180.56 & 139.9 & 161.82 & & $1.962\cdot 10^{-4}$ & $5.237\cdot 10^{-3}$\\
    
        \texttt{ibm\_geneva} & 24, 25, 26 & 305.17 & 353.86 & 444.63 & &  358.17 & 197.83 & 392.43 & & $9.966\cdot 10^{-4}$ & $7.725\cdot 10^{-3}$\\

        \texttt{ibm\_peekskill} & 22, 24, 25 & 342.00 & 269.12 & 161.79 & & 335.64 & 333.85 & 418.65 & & $1.192\cdot 10^{-4}$ & $4.417\cdot 10^{-3}$\\
        
        \hline
        \texttt{ibmq\_lima} & 0, 1, 3 ,4 & 106.58 & 59.93 & 106.47 & 23.72 &  180.56 & 139.9 & 99.2 & 30.62 &  $6.403\cdot 10^{-4}$ & $1.602\cdot 10^{-2}$\\
    \end{tabular}
    \caption{Excerpts from the calibration of the superconducting qubits employed by IBM around the time of the tetraquark experiments one and two (upper three devices), and the last experiment examining the pentaquark (lowest row).}
    \label{tab:calib_exp12}
\end{table*}
\begin{figure}[t]
\includegraphics[width=0.48\textwidth]{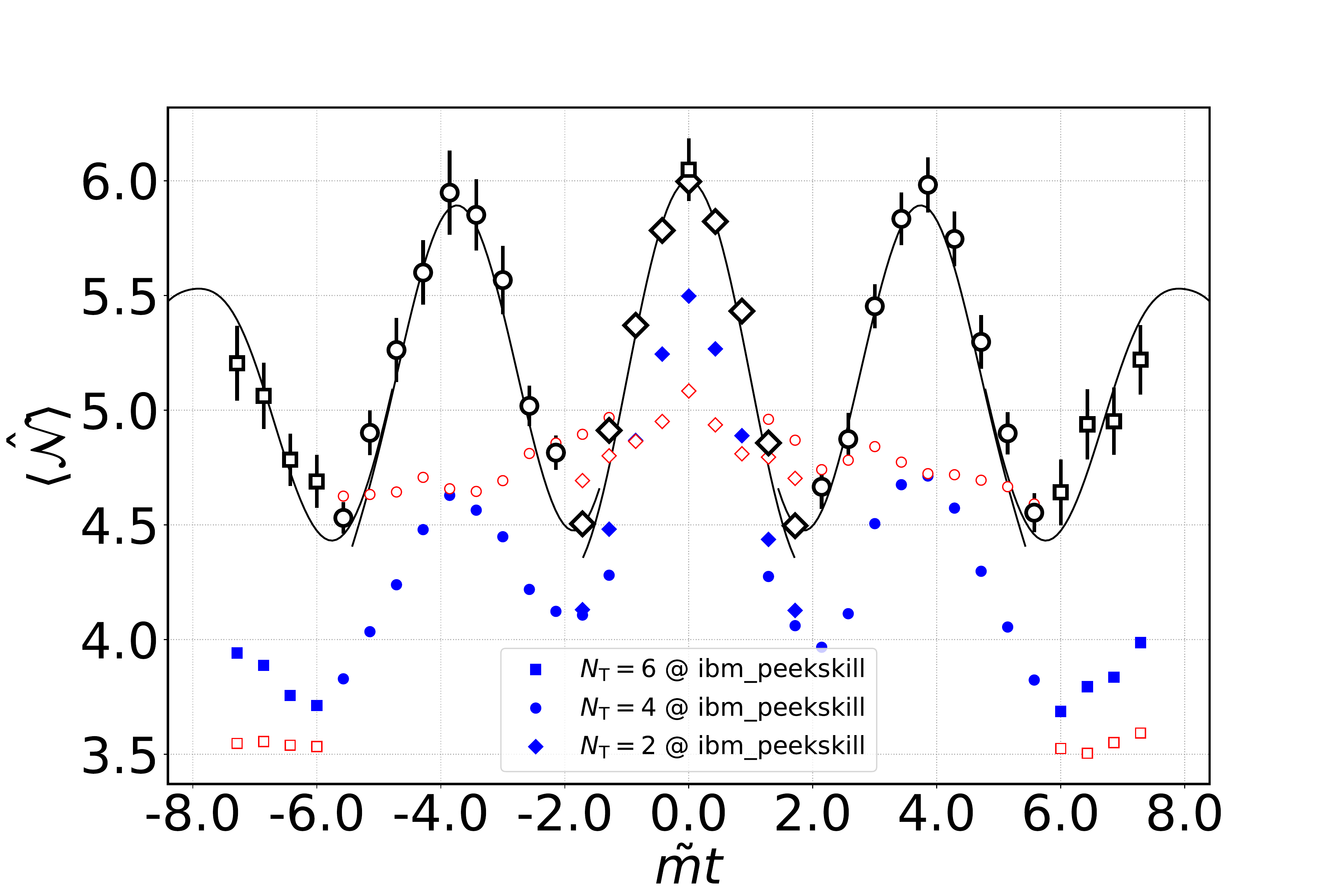} \\ \includegraphics[width=0.48\textwidth]{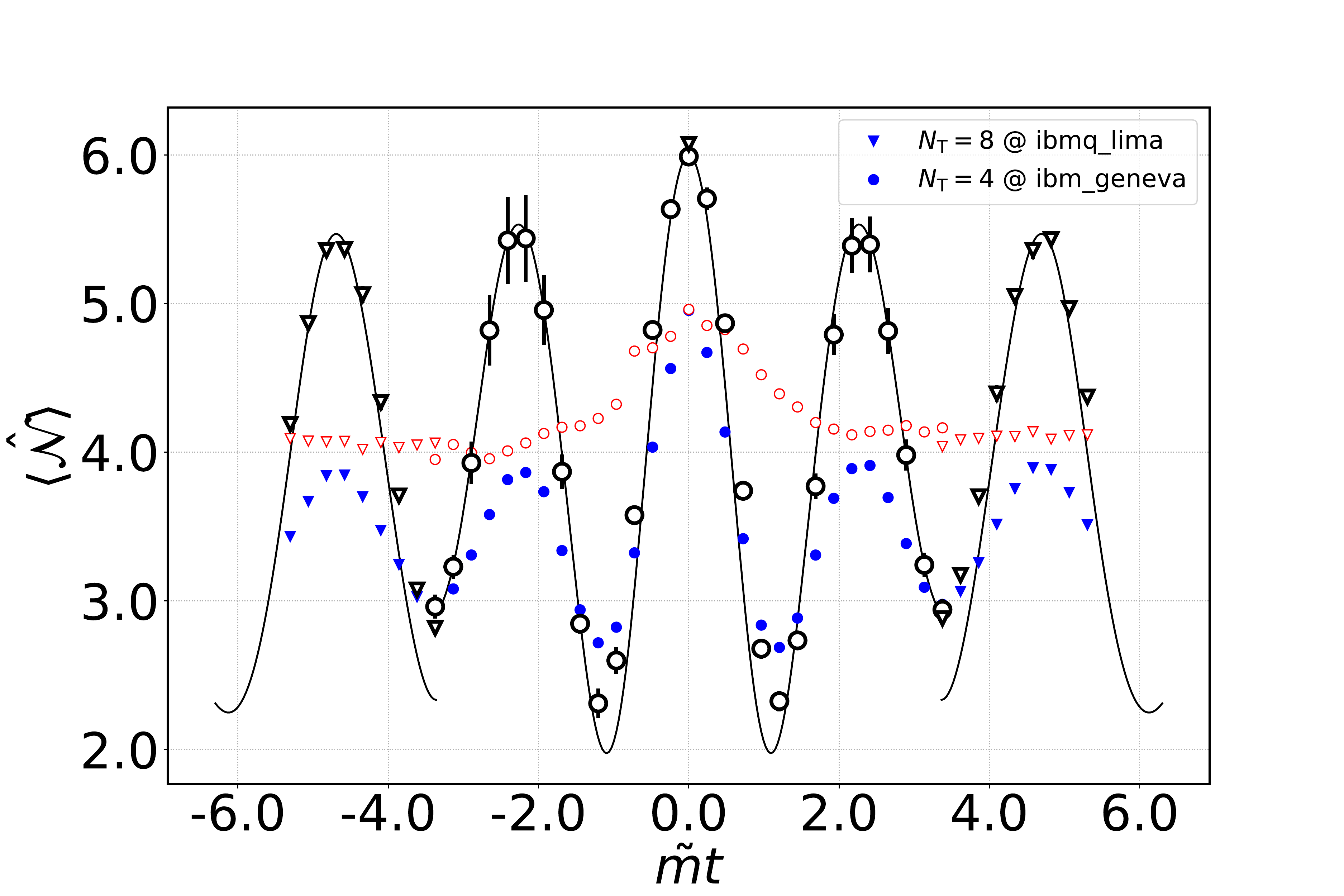}
\caption{Experimental results for $\tilde{m} = 1.2$ and $x=0.8$ (top panel). The mitigated data (large markers with error bars) have already been shown in the main text. The blue (solid) markers correspond to the physical runs, while the red markers (white faces) are the result of the mitigation runs. The shape of the markers denotes the specific IBM machine that has been employed, (see main text). The lower panel shows the equivalent data for the case $\tilde{m}=0.45$.}
\label{fig:snake_SI_plot}
\end{figure}
\begin{figure}[t]
\includegraphics[width=0.48\textwidth]{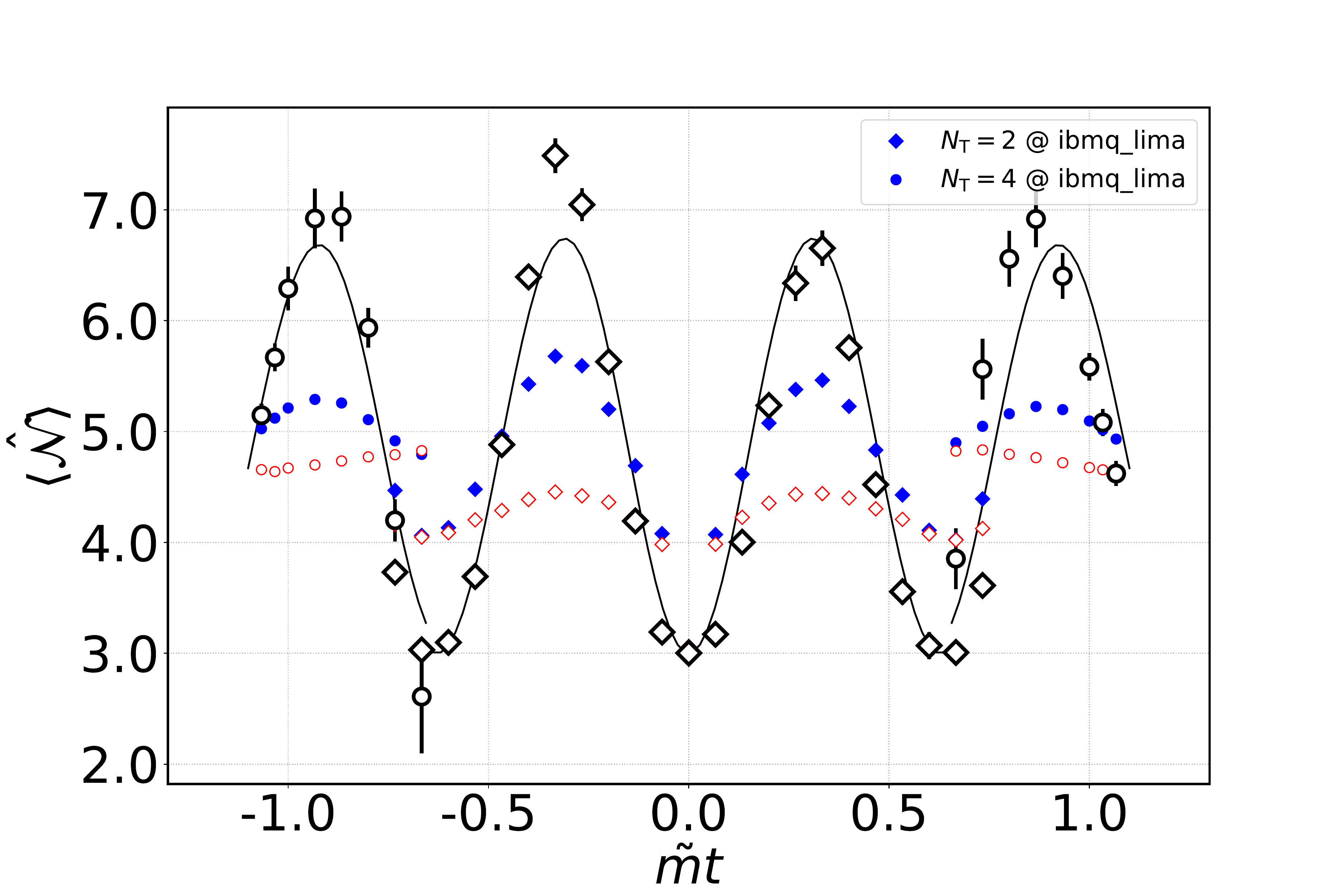}
\caption{Experimental results for the pentaquark, for $\tilde{m} = 0.1$ and $x=3.0$. As in Fig. \ref{fig:snake_SI_plot}, the blue points mark the physical runs, while the red markes with white face correspond to the mitigation runs. As before, the shape of the markers denotes the number Trotter steps.}
\label{fig:pentaquark_SI_plot}
\end{figure}
\subsection{Quantum hardware and error mitigation} \label{appendix:error_mitigation}
In the main text we noted that the experiments have been carried out on various quantum computers in the IBM line-up. Table \ref{tab:calib_exp12} contains excerpts of the calibration taken by IBM during the time of our calculations. Note that $CNOT$ gates take around $300$ to $500\, \mathrm{ns}$, hence for 80 $CNOT$ gates the execution length is on the order of $40\,\mathrm{\mu s}$, which is faster than any of the measured coherence times for the qubits used in the experiments. 

As described in the main text, we apply the randomized compiling techniques as described in \cite{rahman2022} to reduce errors introduced by imperfect $CNOT$ gates. Here, we perform an average over different executions of the same base circuit, where each $CNOT$ operation $\hat{U}_{cx}$ is replaced by $\hat{U}_r\hat{U}_{cx}\hat{U}_r^\dagger$, where $\hat{U}_r$ is a basis transformation that, in an ideal setting, would not alter the action of the $CNOT$ gate. The $\hat{U}_r$ are randomly chosen from a list of 16 possibilities (see \cite{rahman2022} for the full list). The averaging then transforms any coherent gate errors into incoherent ones, which can be equivalently understood as the random unitary model of quantum decoherence, where decoherence is introduced via a classical, randomly fluctuating field. 
\begin{table*}[]
    \centering
    \begin{tabular}{c||cc|cc}
         & \multicolumn{2}{c}{$\tilde{m}=1.2$} & \multicolumn{2}{c}{$\tilde{m}=0.45$} \\ \hline
         & point estimate & HDI & point estimate & HDI \\ \hline \hline
        $A_1$ & 0.689 & $[0.582,\, 0.752]$ & $0.622$ & $[0.252,\, 1.560]$\\
        $A_2$ & -0.0135 & $[-0.014,\, 0.122]$ & $1.182$ & $[0.352,\, 1.606]$\\
        $\omega_1/(2\pi)$ & 0.262 & $[0.254,\, 0.267]$ & $0.482$ & $[0.409,\, 0.545]$\\
        $\omega_2/(2\pi)$ & 0.338 & $[0.000,\, 0.712]$ & $0.427$ & $[0.297,\, 0.502]$\\
        $\phi_1$ & -0.045 & $[-0.102,\, 0.020]$ & $0.009$ & $[-0.114,\, 0.186]$\\
        $\phi_2$ & -0.549& $[-3.425,\, 2.724]$ & $-0.025$ & $[-0.083,\, 0.032]$\\
        $\xi$ & 5.243 & $[5.127,\, 5.304]$ & $4.220$ & $[4.135,\, 4.299]$\\
        $\sigma_2$ & 0.056 & $[0.026,\, 0.091]$ & &\\
        $\sigma_4$ & 0.087 & $[0.047,\, 0.130]$ & $0.226$ & $[0.083,\, 0.315]$\\
        $\sigma_6$ & 0.340 & $[0.208,\, 0.461]$\\
        $\sigma_8$ & & & $0.145$ & $[0.038,\, 0.432]$
    \end{tabular}
    \caption{Bayesian point estimates and HDI for the two tetraquark experiments, where $x=0.8$ and $\tilde{m}=1.2$ or $\tilde{m}=0.45$ respectively.}
    \label{tab:bayes_estimates}
\end{table*}
We perform two kinds of Trotter evolution circuits on the hardware. We first perform a physical run, which aims to evolve the state up to the final time $t=t_f$, and then a second mitigation circuit evolves the state up to $t=t^\prime$ and back to $t=0$. For Trotter steps where $N_\mathrm{T}/2$ is even, we have $t^\prime = t_f/2$. Since our circuit always applies two Trotter steps at once, for $N_\mathrm{T} = 2$ ($6$) we have $t^\prime=t_f$ ($\frac{4}{3}t_f$) and the mitigation circuit contains more $CNOT$ gates than the physical circuit, which we have to correct for. During the experiment, we alternate between physical and mitigation circuits to avoid any biases that could be caused by slow drifts in the experimental setting. We then correct the expectation value of an observable $\hat{O}$ by comparing its measured value for the physical circuits (index phys, meas) and the measured value for the mitigation circuits (index mitig, meas) as
\begin{equation}
    \langle \hat{O} \rangle_\mathrm{phys,\,true} = \left(\frac{\langle \hat{O} \rangle_\mathrm{mitig,\,true}}{\langle \hat{O} \rangle_\mathrm{mitig,\,meas}}\right)^{\kappa} 
    \langle \hat{O} \rangle_\mathrm{phys,\,meas},
\end{equation}
where $\kappa$ is the ratio of the number of $CNOT$ gates in the physical and the mitigation circuits. Note that $\langle \hat{O} \rangle_\mathrm{mitig,\,true}$ is known, since it is given through the state at $t=0$ which is the initial state.

In Fig.~\ref{fig:snake_SI_plot} we show the collected data that have been used to produce Fig. 3 of the main text. The physical runs are shown in blue, while the corresponding mitigation runs are shown in red, with the symbols according to the mitigated data.
Note that for $N_\mathrm{T}\in \lbrace 2,6\rbrace$ we have $\kappa=\frac{1}{2}$ and $\kappa = 1$ otherwise. 
Further, the mitigation has been performed on the observable \mbox{$\hat{H}_m^{(3)} - 3 = - (\hat{Z}_1+\hat{Z}_2+\hat{Z}_3)$}. 
For Fig.~\ref{fig:pentaquark_SI_plot} we have $\kappa=1$ and mitigated $\hat{H}_m^{(4)}-3$.

\subsection{Bayesian inference analysis}
\label{app:bayes}
We model the obtained data $D$ as draws from a normal distribution with variance $\sigma$, where the mean $S_K$ is given by a cosine series 
\begin{equation}
    S_K = \sum_i^K A_i \cos\left(\omega_i t + \phi_i\right) + \xi, 
\end{equation}
with uniform priors on the frequencies $\omega_i$ and the amplitudes $A_i$, while the priors for the phases $\phi_i$ and the offset $\xi$ are normal distributions located at zero. The uniform priors are adjusted to include the frequencies that are visible from the data by eye. The prior on $\sigma$ is given by the maximum of $0.3$ and the value of the largest error bar in the data set. Note that each data set stemming from different $N_\mathrm{T}$ is modelled as a separate likelihood while the parameters are shared. 
After obtaining a representation of the posterior distribution for the parameters $\theta = \{A_i, \omega_i, \phi_i, \xi, \sigma\}$, $P\left(\theta \vert D\right)$, via Monte Carlo sampling from Bayes' rule \cite{kruschke2014}, we sample the posterior predictive distribution $P(D^\prime\vert D) = \int_\Theta P(\Theta=\theta\vert D)P(D^\prime \vert \Theta = \theta)$, where $\Theta$ denotes the collective random variables for all parameters. The values for the point estimates and HDI for all parameters can be found in the Table \ref{tab:bayes_estimates}. Note that in both cases we chose $K=2$. For $\tilde{m}=1.2$, the low amplitude of the second frequency demonstrates well that the dynamics are indeed dominated by the transition of the baryon-antibaryon and the tetra quark state, while we find two competing frequencies in the case $\tilde{m}=0.45$. Furthermore, we note that in the latter case, the highest density interval (HDI) of the frequencies are not overlapping, while the HDI for $\omega_2$ in the $\tilde{m}=1.2$ case spans a wide range, indicating that the single frequency $\omega_1$ is mostly enough to describe the whole recorded signal.



\bibliography{SU3_biblio}{}

\begin{thebibliography}{66}%
\makeatletter
\providecommand \@ifxundefined [1]{%
 \@ifx{#1\undefined}
}%
\providecommand \@ifnum [1]{%
 \ifnum #1\expandafter \@firstoftwo
 \else \expandafter \@secondoftwo
 \fi
}%
\providecommand \@ifx [1]{%
 \ifx #1\expandafter \@firstoftwo
 \else \expandafter \@secondoftwo
 \fi
}%
\providecommand \natexlab [1]{#1}%
\providecommand \enquote  [1]{``#1''}%
\providecommand \bibnamefont  [1]{#1}%
\providecommand \bibfnamefont [1]{#1}%
\providecommand \citenamefont [1]{#1}%
\providecommand \href@noop [0]{\@secondoftwo}%
\providecommand \href [0]{\begingroup \@sanitize@url \@href}%
\providecommand \@href[1]{\@@startlink{#1}\@@href}%
\providecommand \@@href[1]{\endgroup#1\@@endlink}%
\providecommand \@sanitize@url [0]{\catcode `\\12\catcode `\$12\catcode
  `\&12\catcode `\#12\catcode `\^12\catcode `\_12\catcode `\%12\relax}%
\providecommand \@@startlink[1]{}%
\providecommand \@@endlink[0]{}%
\providecommand \url  [0]{\begingroup\@sanitize@url \@url }%
\providecommand \@url [1]{\endgroup\@href {#1}{\urlprefix }}%
\providecommand \urlprefix  [0]{URL }%
\providecommand \Eprint [0]{\href }%
\providecommand \doibase [0]{https://doi.org/}%
\providecommand \selectlanguage [0]{\@gobble}%
\providecommand \bibinfo  [0]{\@secondoftwo}%
\providecommand \bibfield  [0]{\@secondoftwo}%
\providecommand \translation [1]{[#1]}%
\providecommand \BibitemOpen [0]{}%
\providecommand \bibitemStop [0]{}%
\providecommand \bibitemNoStop [0]{.\EOS\space}%
\providecommand \EOS [0]{\spacefactor3000\relax}%
\providecommand \BibitemShut  [1]{\csname bibitem#1\endcsname}%
\let\auto@bib@innerbib\@empty
\bibitem [{\citenamefont {Chen}\ \emph {et~al.}(2022)\citenamefont {Chen},
  \citenamefont {Chen}, \citenamefont {Liu}, \citenamefont {Liu},\ and\
  \citenamefont {Zhu}}]{Chen:2022asf}%
  \BibitemOpen
  \bibfield  {author} {\bibinfo {author} {\bibfnamefont {H.-X.}\ \bibnamefont
  {Chen}}, \bibinfo {author} {\bibfnamefont {W.}~\bibnamefont {Chen}}, \bibinfo
  {author} {\bibfnamefont {X.}~\bibnamefont {Liu}}, \bibinfo {author}
  {\bibfnamefont {Y.-R.}\ \bibnamefont {Liu}},\ and\ \bibinfo {author}
  {\bibfnamefont {S.-L.}\ \bibnamefont {Zhu}},\ }\bibfield  {title} {\bibinfo
  {title} {An updated review of the new hadron states},\ }\href@noop {}
  {\bibfield  {journal} {\bibinfo  {journal} {arXiv preprint arXiv:2204.02649}\
  } (\bibinfo {year} {2022})}\BibitemShut {NoStop}%
\bibitem [{\citenamefont {Ba{\~n}uls}\ \emph {et~al.}(2020)\citenamefont
  {Ba{\~n}uls}, \citenamefont {Blatt}, \citenamefont {Catani}, \citenamefont
  {Celi}, \citenamefont {Cirac}, \citenamefont {Dalmonte}, \citenamefont
  {Fallani}, \citenamefont {Jansen}, \citenamefont {Lewenstein}, \citenamefont
  {Montangero}, \citenamefont {Muschik}, \citenamefont {Reznik}, \citenamefont
  {Rico}, \citenamefont {Tagliacozzo}, \citenamefont {Van~Acoleyen},
  \citenamefont {Verstraete}, \citenamefont {Wiese}, \citenamefont {Wingate},
  \citenamefont {Zakrzewski},\ and\ \citenamefont
  {Zoller}}]{banuls_simulating_2020}%
  \BibitemOpen
  \bibfield  {author} {\bibinfo {author} {\bibfnamefont {M.~C.}\ \bibnamefont
  {Ba{\~n}uls}}, \bibinfo {author} {\bibfnamefont {R.}~\bibnamefont {Blatt}},
  \bibinfo {author} {\bibfnamefont {J.}~\bibnamefont {Catani}}, \bibinfo
  {author} {\bibfnamefont {A.}~\bibnamefont {Celi}}, \bibinfo {author}
  {\bibfnamefont {J.~I.}\ \bibnamefont {Cirac}}, \bibinfo {author}
  {\bibfnamefont {M.}~\bibnamefont {Dalmonte}}, \bibinfo {author}
  {\bibfnamefont {L.}~\bibnamefont {Fallani}}, \bibinfo {author} {\bibfnamefont
  {K.}~\bibnamefont {Jansen}}, \bibinfo {author} {\bibfnamefont
  {M.}~\bibnamefont {Lewenstein}}, \bibinfo {author} {\bibfnamefont
  {S.}~\bibnamefont {Montangero}}, \bibinfo {author} {\bibfnamefont {C.~A.}\
  \bibnamefont {Muschik}}, \bibinfo {author} {\bibfnamefont {B.}~\bibnamefont
  {Reznik}}, \bibinfo {author} {\bibfnamefont {E.}~\bibnamefont {Rico}},
  \bibinfo {author} {\bibfnamefont {L.}~\bibnamefont {Tagliacozzo}}, \bibinfo
  {author} {\bibfnamefont {K.}~\bibnamefont {Van~Acoleyen}}, \bibinfo {author}
  {\bibfnamefont {F.}~\bibnamefont {Verstraete}}, \bibinfo {author}
  {\bibfnamefont {U.-J.}\ \bibnamefont {Wiese}}, \bibinfo {author}
  {\bibfnamefont {M.}~\bibnamefont {Wingate}}, \bibinfo {author} {\bibfnamefont
  {J.}~\bibnamefont {Zakrzewski}},\ and\ \bibinfo {author} {\bibfnamefont
  {P.}~\bibnamefont {Zoller}},\ }\bibfield  {title} {\bibinfo {title}
  {Simulating lattice gauge theories within quantum technologies},\ }\href
  {https://doi.org/10.1140/epjd/e2020-100571-8} {\bibfield  {journal} {\bibinfo
   {journal} {Eur. Phys. J. D}\ }\textbf {\bibinfo {volume} {74}},\ \bibinfo
  {pages} {165} (\bibinfo {year} {2020})}\BibitemShut {NoStop}%
\bibitem [{\citenamefont {Davoudi}\ \emph {et~al.}(2022)\citenamefont
  {Davoudi}, \citenamefont {Bauer}, \citenamefont {Balantekin}, \citenamefont
  {Bhattacharya}, \citenamefont {Carena}, \citenamefont {de~Jong},
  \citenamefont {Draper}, \citenamefont {El-Khadra}, \citenamefont {Gemelke},
  \citenamefont {Hanada}, \citenamefont {Kharzeev} \emph
  {et~al.}}]{bauer_quantum_2022}%
  \BibitemOpen
  \bibfield  {author} {\bibinfo {author} {\bibfnamefont {Z.}~\bibnamefont
  {Davoudi}}, \bibinfo {author} {\bibfnamefont {C.~W.}\ \bibnamefont {Bauer}},
  \bibinfo {author} {\bibfnamefont {A.}~\bibnamefont {Balantekin}}, \bibinfo
  {author} {\bibfnamefont {T.}~\bibnamefont {Bhattacharya}}, \bibinfo {author}
  {\bibfnamefont {M.}~\bibnamefont {Carena}}, \bibinfo {author} {\bibfnamefont
  {W.~A.}\ \bibnamefont {de~Jong}}, \bibinfo {author} {\bibfnamefont
  {P.}~\bibnamefont {Draper}}, \bibinfo {author} {\bibfnamefont
  {A.}~\bibnamefont {El-Khadra}}, \bibinfo {author} {\bibfnamefont
  {N.}~\bibnamefont {Gemelke}}, \bibinfo {author} {\bibfnamefont
  {M.}~\bibnamefont {Hanada}}, \bibinfo {author} {\bibfnamefont
  {D.}~\bibnamefont {Kharzeev}}, \emph {et~al.},\ }\bibfield  {title} {\bibinfo
  {title} {Quantum simulation for high energy physics},\ }\href@noop {}
  {\bibfield  {journal} {\bibinfo  {journal} {arXiv preprint arXiv:2204.03381}\
  } (\bibinfo {year} {2022})}\BibitemShut {NoStop}%
\bibitem [{\citenamefont {Dalmonte}\ and\ \citenamefont
  {Montangero}(2016)}]{dalmonte2016lattice}%
  \BibitemOpen
  \bibfield  {author} {\bibinfo {author} {\bibfnamefont {M.}~\bibnamefont
  {Dalmonte}}\ and\ \bibinfo {author} {\bibfnamefont {S.}~\bibnamefont
  {Montangero}},\ }\bibfield  {title} {\bibinfo {title} {Lattice gauge theory
  simulations in the quantum information era},\ }\href@noop {} {\bibfield
  {journal} {\bibinfo  {journal} {Contemporary Physics}\ }\textbf {\bibinfo
  {volume} {57}},\ \bibinfo {pages} {388} (\bibinfo {year} {2016})}\BibitemShut
  {NoStop}%
\bibitem [{\citenamefont {Silvi}\ \emph {et~al.}(2019)\citenamefont {Silvi},
  \citenamefont {Sauer}, \citenamefont {Tschirsich},\ and\ \citenamefont
  {Montangero}}]{silvi_tensor_2019}%
  \BibitemOpen
  \bibfield  {author} {\bibinfo {author} {\bibfnamefont {P.}~\bibnamefont
  {Silvi}}, \bibinfo {author} {\bibfnamefont {Y.}~\bibnamefont {Sauer}},
  \bibinfo {author} {\bibfnamefont {F.}~\bibnamefont {Tschirsich}},\ and\
  \bibinfo {author} {\bibfnamefont {S.}~\bibnamefont {Montangero}},\ }\bibfield
   {title} {\bibinfo {title} {Tensor network simulation of an {{SU}}(3) lattice
  gauge theory in {{1D}}},\ }\href
  {https://doi.org/10.1103/PhysRevD.100.074512} {\bibfield  {journal} {\bibinfo
   {journal} {Phys. Rev. D}\ }\textbf {\bibinfo {volume} {100}},\ \bibinfo
  {pages} {074512} (\bibinfo {year} {2019})}\BibitemShut {NoStop}%
\bibitem [{\citenamefont {Martinez}\ \emph {et~al.}(2016)\citenamefont
  {Martinez}, \citenamefont {Muschik}, \citenamefont {Schindler}, \citenamefont
  {Nigg}, \citenamefont {Erhard}, \citenamefont {Heyl}, \citenamefont {Hauke},
  \citenamefont {Dalmonte}, \citenamefont {Monz}, \citenamefont {Zoller},\ and\
  \citenamefont {Blatt}}]{martinez_real-time_2016}%
  \BibitemOpen
  \bibfield  {author} {\bibinfo {author} {\bibfnamefont {E.~A.}\ \bibnamefont
  {Martinez}}, \bibinfo {author} {\bibfnamefont {C.~A.}\ \bibnamefont
  {Muschik}}, \bibinfo {author} {\bibfnamefont {P.}~\bibnamefont {Schindler}},
  \bibinfo {author} {\bibfnamefont {D.}~\bibnamefont {Nigg}}, \bibinfo {author}
  {\bibfnamefont {A.}~\bibnamefont {Erhard}}, \bibinfo {author} {\bibfnamefont
  {M.}~\bibnamefont {Heyl}}, \bibinfo {author} {\bibfnamefont {P.}~\bibnamefont
  {Hauke}}, \bibinfo {author} {\bibfnamefont {M.}~\bibnamefont {Dalmonte}},
  \bibinfo {author} {\bibfnamefont {T.}~\bibnamefont {Monz}}, \bibinfo {author}
  {\bibfnamefont {P.}~\bibnamefont {Zoller}},\ and\ \bibinfo {author}
  {\bibfnamefont {R.}~\bibnamefont {Blatt}},\ }\bibfield  {title} {\bibinfo
  {title} {Real-time dynamics of lattice gauge theories with a few-qubit
  quantum computer},\ }\href {https://doi.org/10.1038/nature18318} {\bibfield
  {journal} {\bibinfo  {journal} {Nature}\ }\textbf {\bibinfo {volume} {534}},\
  \bibinfo {pages} {516} (\bibinfo {year} {2016})}\BibitemShut {NoStop}%
\bibitem [{\citenamefont {Klco}\ \emph {et~al.}(2018)\citenamefont {Klco},
  \citenamefont {Dumitrescu}, \citenamefont {McCaskey}, \citenamefont {Morris},
  \citenamefont {Pooser}, \citenamefont {Sanz}, \citenamefont {Solano},
  \citenamefont {Lougovski},\ and\ \citenamefont
  {Savage}}]{klco_quantum-classical_2018}%
  \BibitemOpen
  \bibfield  {author} {\bibinfo {author} {\bibfnamefont {N.}~\bibnamefont
  {Klco}}, \bibinfo {author} {\bibfnamefont {E.~F.}\ \bibnamefont
  {Dumitrescu}}, \bibinfo {author} {\bibfnamefont {A.~J.}\ \bibnamefont
  {McCaskey}}, \bibinfo {author} {\bibfnamefont {T.~D.}\ \bibnamefont
  {Morris}}, \bibinfo {author} {\bibfnamefont {R.~C.}\ \bibnamefont {Pooser}},
  \bibinfo {author} {\bibfnamefont {M.}~\bibnamefont {Sanz}}, \bibinfo {author}
  {\bibfnamefont {E.}~\bibnamefont {Solano}}, \bibinfo {author} {\bibfnamefont
  {P.}~\bibnamefont {Lougovski}},\ and\ \bibinfo {author} {\bibfnamefont
  {M.~J.}\ \bibnamefont {Savage}},\ }\bibfield  {title} {\bibinfo {title}
  {Quantum-{{Classical Computation}} of {{Schwinger Model Dynamics}} using
  {{Quantum Computers}}},\ }\href {https://doi.org/10.1103/PhysRevA.98.032331}
  {\bibfield  {journal} {\bibinfo  {journal} {Phys. Rev. A}\ }\textbf {\bibinfo
  {volume} {98}},\ \bibinfo {pages} {032331} (\bibinfo {year}
  {2018})}\BibitemShut {NoStop}%
\bibitem [{\citenamefont {Kokail}\ \emph {et~al.}(2019)\citenamefont {Kokail},
  \citenamefont {Maier}, \citenamefont {van Bijnen}, \citenamefont {Brydges},
  \citenamefont {Joshi}, \citenamefont {Jurcevic}, \citenamefont {Muschik},
  \citenamefont {Silvi}, \citenamefont {Blatt}, \citenamefont {Roos},\ and\
  \citenamefont {Zoller}}]{kokail_self-verifying_2019}%
  \BibitemOpen
  \bibfield  {author} {\bibinfo {author} {\bibfnamefont {C.}~\bibnamefont
  {Kokail}}, \bibinfo {author} {\bibfnamefont {C.}~\bibnamefont {Maier}},
  \bibinfo {author} {\bibfnamefont {R.}~\bibnamefont {van Bijnen}}, \bibinfo
  {author} {\bibfnamefont {T.}~\bibnamefont {Brydges}}, \bibinfo {author}
  {\bibfnamefont {M.~K.}\ \bibnamefont {Joshi}}, \bibinfo {author}
  {\bibfnamefont {P.}~\bibnamefont {Jurcevic}}, \bibinfo {author}
  {\bibfnamefont {C.~A.}\ \bibnamefont {Muschik}}, \bibinfo {author}
  {\bibfnamefont {P.}~\bibnamefont {Silvi}}, \bibinfo {author} {\bibfnamefont
  {R.}~\bibnamefont {Blatt}}, \bibinfo {author} {\bibfnamefont {C.~F.}\
  \bibnamefont {Roos}},\ and\ \bibinfo {author} {\bibfnamefont
  {P.}~\bibnamefont {Zoller}},\ }\bibfield  {title} {\bibinfo {title}
  {Self-verifying variational quantum simulation of lattice models},\ }\href
  {https://doi.org/10.1038/s41586-019-1177-4} {\bibfield  {journal} {\bibinfo
  {journal} {Nature}\ }\textbf {\bibinfo {volume} {569}},\ \bibinfo {pages}
  {355} (\bibinfo {year} {2019})}\BibitemShut {NoStop}%
\bibitem [{\citenamefont {Lu}\ \emph {et~al.}(2019)\citenamefont {Lu},
  \citenamefont {Klco}, \citenamefont {Lukens}, \citenamefont {Morris},
  \citenamefont {Bansal}, \citenamefont {Ekstr{\"o}m}, \citenamefont {Hagen},
  \citenamefont {Papenbrock}, \citenamefont {Weiner}, \citenamefont {Savage},\
  and\ \citenamefont {Lougovski}}]{lu_simulations_2019}%
  \BibitemOpen
  \bibfield  {author} {\bibinfo {author} {\bibfnamefont {H.-H.}\ \bibnamefont
  {Lu}}, \bibinfo {author} {\bibfnamefont {N.}~\bibnamefont {Klco}}, \bibinfo
  {author} {\bibfnamefont {J.~M.}\ \bibnamefont {Lukens}}, \bibinfo {author}
  {\bibfnamefont {T.~D.}\ \bibnamefont {Morris}}, \bibinfo {author}
  {\bibfnamefont {A.}~\bibnamefont {Bansal}}, \bibinfo {author} {\bibfnamefont
  {A.}~\bibnamefont {Ekstr{\"o}m}}, \bibinfo {author} {\bibfnamefont
  {G.}~\bibnamefont {Hagen}}, \bibinfo {author} {\bibfnamefont
  {T.}~\bibnamefont {Papenbrock}}, \bibinfo {author} {\bibfnamefont {A.~M.}\
  \bibnamefont {Weiner}}, \bibinfo {author} {\bibfnamefont {M.~J.}\
  \bibnamefont {Savage}},\ and\ \bibinfo {author} {\bibfnamefont
  {P.}~\bibnamefont {Lougovski}},\ }\bibfield  {title} {\bibinfo {title}
  {Simulations of subatomic many-body physics on a quantum frequency
  processor},\ }\href {https://doi.org/10.1103/PhysRevA.100.012320} {\bibfield
  {journal} {\bibinfo  {journal} {Phys. Rev. A}\ }\textbf {\bibinfo {volume}
  {100}},\ \bibinfo {pages} {012320} (\bibinfo {year} {2019})}\BibitemShut
  {NoStop}%
\bibitem [{\citenamefont {Mil}\ \emph {et~al.}(2020)\citenamefont {Mil},
  \citenamefont {Zache}, \citenamefont {Hegde}, \citenamefont {Xia},
  \citenamefont {Bhatt}, \citenamefont {Oberthaler}, \citenamefont {Hauke},
  \citenamefont {Berges},\ and\ \citenamefont
  {Jendrzejewski}}]{mil_scalable_2020}%
  \BibitemOpen
  \bibfield  {author} {\bibinfo {author} {\bibfnamefont {A.}~\bibnamefont
  {Mil}}, \bibinfo {author} {\bibfnamefont {T.~V.}\ \bibnamefont {Zache}},
  \bibinfo {author} {\bibfnamefont {A.}~\bibnamefont {Hegde}}, \bibinfo
  {author} {\bibfnamefont {A.}~\bibnamefont {Xia}}, \bibinfo {author}
  {\bibfnamefont {R.~P.}\ \bibnamefont {Bhatt}}, \bibinfo {author}
  {\bibfnamefont {M.~K.}\ \bibnamefont {Oberthaler}}, \bibinfo {author}
  {\bibfnamefont {P.}~\bibnamefont {Hauke}}, \bibinfo {author} {\bibfnamefont
  {J.}~\bibnamefont {Berges}},\ and\ \bibinfo {author} {\bibfnamefont
  {F.}~\bibnamefont {Jendrzejewski}},\ }\bibfield  {title} {\bibinfo {title} {A
  scalable realization of local {{U}}(1) gauge invariance in cold atomic
  mixtures},\ }\href {https://doi.org/10.1126/science.aaz5312} {\bibfield
  {journal} {\bibinfo  {journal} {Science}\ }\textbf {\bibinfo {volume}
  {367}},\ \bibinfo {pages} {1128} (\bibinfo {year} {2020})}\BibitemShut
  {NoStop}%
\bibitem [{\citenamefont {Surace}\ \emph {et~al.}(2020)\citenamefont {Surace},
  \citenamefont {Mazza}, \citenamefont {Giudici}, \citenamefont {Lerose},
  \citenamefont {Gambassi},\ and\ \citenamefont
  {Dalmonte}}]{surace_lattice_2020}%
  \BibitemOpen
  \bibfield  {author} {\bibinfo {author} {\bibfnamefont {F.~M.}\ \bibnamefont
  {Surace}}, \bibinfo {author} {\bibfnamefont {P.~P.}\ \bibnamefont {Mazza}},
  \bibinfo {author} {\bibfnamefont {G.}~\bibnamefont {Giudici}}, \bibinfo
  {author} {\bibfnamefont {A.}~\bibnamefont {Lerose}}, \bibinfo {author}
  {\bibfnamefont {A.}~\bibnamefont {Gambassi}},\ and\ \bibinfo {author}
  {\bibfnamefont {M.}~\bibnamefont {Dalmonte}},\ }\bibfield  {title} {\bibinfo
  {title} {Lattice {{Gauge Theories}} and {{String Dynamics}} in {{Rydberg Atom
  Quantum Simulators}}},\ }\href {https://doi.org/10.1103/PhysRevX.10.021041}
  {\bibfield  {journal} {\bibinfo  {journal} {Phys. Rev. X}\ }\textbf {\bibinfo
  {volume} {10}},\ \bibinfo {pages} {021041} (\bibinfo {year}
  {2020})}\BibitemShut {NoStop}%
\bibitem [{\citenamefont {Yang}\ \emph {et~al.}(2020)\citenamefont {Yang},
  \citenamefont {Sun}, \citenamefont {Ott}, \citenamefont {Wang}, \citenamefont
  {Zache}, \citenamefont {Halimeh}, \citenamefont {Yuan}, \citenamefont
  {Hauke},\ and\ \citenamefont {Pan}}]{yang_observation_2020}%
  \BibitemOpen
  \bibfield  {author} {\bibinfo {author} {\bibfnamefont {B.}~\bibnamefont
  {Yang}}, \bibinfo {author} {\bibfnamefont {H.}~\bibnamefont {Sun}}, \bibinfo
  {author} {\bibfnamefont {R.}~\bibnamefont {Ott}}, \bibinfo {author}
  {\bibfnamefont {H.-Y.}\ \bibnamefont {Wang}}, \bibinfo {author}
  {\bibfnamefont {T.~V.}\ \bibnamefont {Zache}}, \bibinfo {author}
  {\bibfnamefont {J.~C.}\ \bibnamefont {Halimeh}}, \bibinfo {author}
  {\bibfnamefont {Z.-S.}\ \bibnamefont {Yuan}}, \bibinfo {author}
  {\bibfnamefont {P.}~\bibnamefont {Hauke}},\ and\ \bibinfo {author}
  {\bibfnamefont {J.-W.}\ \bibnamefont {Pan}},\ }\bibfield  {title} {\bibinfo
  {title} {Observation of gauge invariance in a 71-site
  {{Bose}}\textendash{{Hubbard}} quantum simulator},\ }\href
  {https://doi.org/10.1038/s41586-020-2910-8} {\bibfield  {journal} {\bibinfo
  {journal} {Nature}\ }\textbf {\bibinfo {volume} {587}},\ \bibinfo {pages}
  {392} (\bibinfo {year} {2020})}\BibitemShut {NoStop}%
\bibitem [{\citenamefont {Zhou}\ \emph {et~al.}(2021)\citenamefont {Zhou},
  \citenamefont {Su}, \citenamefont {Halimeh}, \citenamefont {Ott},
  \citenamefont {Sun}, \citenamefont {Hauke}, \citenamefont {Yang},
  \citenamefont {Yuan}, \citenamefont {Berges},\ and\ \citenamefont
  {Pan}}]{zhou_thermalization_2021}%
  \BibitemOpen
  \bibfield  {author} {\bibinfo {author} {\bibfnamefont {Z.}~\bibnamefont
  {Zhou}}, \bibinfo {author} {\bibfnamefont {G.}~\bibnamefont {Su}}, \bibinfo
  {author} {\bibfnamefont {J.}~\bibnamefont {Halimeh}}, \bibinfo {author}
  {\bibfnamefont {R.}~\bibnamefont {Ott}}, \bibinfo {author} {\bibfnamefont
  {H.}~\bibnamefont {Sun}}, \bibinfo {author} {\bibfnamefont {P.}~\bibnamefont
  {Hauke}}, \bibinfo {author} {\bibfnamefont {B.}~\bibnamefont {Yang}},
  \bibinfo {author} {\bibfnamefont {Z.}~\bibnamefont {Yuan}}, \bibinfo {author}
  {\bibfnamefont {J.}~\bibnamefont {Berges}},\ and\ \bibinfo {author}
  {\bibfnamefont {J.}~\bibnamefont {Pan}},\ }\bibfield  {title} {\bibinfo
  {title} {Thermalization dynamics of a gauge theory on a quantum simulator,
  arxiv e-prints},\ }\href@noop {} {\bibfield  {journal} {\bibinfo  {journal}
  {arXiv preprint arXiv:2107.13563}\ } (\bibinfo {year} {2021})}\BibitemShut
  {NoStop}%
\bibitem [{\citenamefont {Nguyen}\ \emph {et~al.}(2022)\citenamefont {Nguyen},
  \citenamefont {Tran}, \citenamefont {Zhu}, \citenamefont {Green},
  \citenamefont {Alderete}, \citenamefont {Davoudi},\ and\ \citenamefont
  {Linke}}]{nguyen_digital_2022}%
  \BibitemOpen
  \bibfield  {author} {\bibinfo {author} {\bibfnamefont {N.~H.}\ \bibnamefont
  {Nguyen}}, \bibinfo {author} {\bibfnamefont {M.~C.}\ \bibnamefont {Tran}},
  \bibinfo {author} {\bibfnamefont {Y.}~\bibnamefont {Zhu}}, \bibinfo {author}
  {\bibfnamefont {A.~M.}\ \bibnamefont {Green}}, \bibinfo {author}
  {\bibfnamefont {C.~H.}\ \bibnamefont {Alderete}}, \bibinfo {author}
  {\bibfnamefont {Z.}~\bibnamefont {Davoudi}},\ and\ \bibinfo {author}
  {\bibfnamefont {N.~M.}\ \bibnamefont {Linke}},\ }\bibfield  {title} {\bibinfo
  {title} {Digital {{Quantum Simulation}} of the {{Schwinger Model}} and
  {{Symmetry Protection}} with {{Trapped Ions}}},\ }\href
  {https://doi.org/10.1103/PRXQuantum.3.020324} {\bibfield  {journal} {\bibinfo
   {journal} {PRX Quantum}\ }\textbf {\bibinfo {volume} {3}},\ \bibinfo {pages}
  {020324} (\bibinfo {year} {2022})}\BibitemShut {NoStop}%
\bibitem [{\citenamefont {Atas}\ \emph {et~al.}(2021)\citenamefont {Atas},
  \citenamefont {Zhang}, \citenamefont {Lewis}, \citenamefont {Jahanpour},
  \citenamefont {Haase},\ and\ \citenamefont {Muschik}}]{atas2021}%
  \BibitemOpen
  \bibfield  {author} {\bibinfo {author} {\bibfnamefont {Y.~Y.}\ \bibnamefont
  {Atas}}, \bibinfo {author} {\bibfnamefont {J.}~\bibnamefont {Zhang}},
  \bibinfo {author} {\bibfnamefont {R.}~\bibnamefont {Lewis}}, \bibinfo
  {author} {\bibfnamefont {A.}~\bibnamefont {Jahanpour}}, \bibinfo {author}
  {\bibfnamefont {J.~F.}\ \bibnamefont {Haase}},\ and\ \bibinfo {author}
  {\bibfnamefont {C.~A.}\ \bibnamefont {Muschik}},\ }\bibfield  {title}
  {\bibinfo {title} {S{U}(2) hadrons on a quantum computer via a variational
  approach},\ }\href@noop {} {\bibfield  {journal} {\bibinfo  {journal} {Nat.
  Commun.}\ }\textbf {\bibinfo {volume} {12}},\ \bibinfo {pages} {6499}
  (\bibinfo {year} {2021})}\BibitemShut {NoStop}%
\bibitem [{\citenamefont {Klco}\ \emph {et~al.}(2020)\citenamefont {Klco},
  \citenamefont {Savage},\ and\ \citenamefont {Stryker}}]{klco_su2_2020}%
  \BibitemOpen
  \bibfield  {author} {\bibinfo {author} {\bibfnamefont {N.}~\bibnamefont
  {Klco}}, \bibinfo {author} {\bibfnamefont {M.~J.}\ \bibnamefont {Savage}},\
  and\ \bibinfo {author} {\bibfnamefont {J.~R.}\ \bibnamefont {Stryker}},\
  }\bibfield  {title} {\bibinfo {title} {{{SU}}(2) non-{{Abelian}} gauge field
  theory in one dimension on digital quantum computers},\ }\href
  {https://doi.org/10.1103/PhysRevD.101.074512} {\bibfield  {journal} {\bibinfo
   {journal} {Phys. Rev. D}\ }\textbf {\bibinfo {volume} {101}},\ \bibinfo
  {pages} {074512} (\bibinfo {year} {2020})}\BibitemShut {NoStop}%
\bibitem [{\citenamefont {A~Rahman}\ \emph {et~al.}(2021)\citenamefont
  {A~Rahman}, \citenamefont {Lewis}, \citenamefont {Mendicelli},\ and\
  \citenamefont {Powell}}]{ARahman:2021ktn}%
  \BibitemOpen
  \bibfield  {author} {\bibinfo {author} {\bibfnamefont {S.}~\bibnamefont
  {A~Rahman}}, \bibinfo {author} {\bibfnamefont {R.}~\bibnamefont {Lewis}},
  \bibinfo {author} {\bibfnamefont {E.}~\bibnamefont {Mendicelli}},\ and\
  \bibinfo {author} {\bibfnamefont {S.}~\bibnamefont {Powell}},\ }\bibfield
  {title} {\bibinfo {title} {{SU(2) lattice gauge theory on a quantum
  annealer}},\ }\href {https://doi.org/10.1103/PhysRevD.104.034501} {\bibfield
  {journal} {\bibinfo  {journal} {Phys. Rev. D}\ }\textbf {\bibinfo {volume}
  {104}},\ \bibinfo {pages} {034501} (\bibinfo {year} {2021})}\BibitemShut
  {NoStop}%
\bibitem [{\citenamefont {Ciavarella}\ \emph {et~al.}(2021)\citenamefont
  {Ciavarella}, \citenamefont {Klco},\ and\ \citenamefont
  {Savage}}]{ciavarella_trailhead_2021}%
  \BibitemOpen
  \bibfield  {author} {\bibinfo {author} {\bibfnamefont {A.}~\bibnamefont
  {Ciavarella}}, \bibinfo {author} {\bibfnamefont {N.}~\bibnamefont {Klco}},\
  and\ \bibinfo {author} {\bibfnamefont {M.~J.}\ \bibnamefont {Savage}},\
  }\bibfield  {title} {\bibinfo {title} {Trailhead for quantum simulation of
  {{SU}}(3) {{Yang-Mills}} lattice gauge theory in the local multiplet basis},\
  }\href {https://doi.org/10.1103/PhysRevD.103.094501} {\bibfield  {journal}
  {\bibinfo  {journal} {Phys. Rev. D}\ }\textbf {\bibinfo {volume} {103}},\
  \bibinfo {pages} {094501} (\bibinfo {year} {2021})}\BibitemShut {NoStop}%
\bibitem [{\citenamefont {Illa}\ and\ \citenamefont
  {Savage}(2022)}]{Illa:2022jqb}%
  \BibitemOpen
  \bibfield  {author} {\bibinfo {author} {\bibfnamefont {M.}~\bibnamefont
  {Illa}}\ and\ \bibinfo {author} {\bibfnamefont {M.~J.}\ \bibnamefont
  {Savage}},\ }\bibfield  {title} {\bibinfo {title} {Basic elements for
  simulations of standard model physics with quantum annealers: Multigrid and
  clock states},\ }\href@noop {} {\bibfield  {journal} {\bibinfo  {journal}
  {arXiv preprint arXiv:2202.12340}\ } (\bibinfo {year} {2022})}\BibitemShut
  {NoStop}%
\bibitem [{\citenamefont {A~Rahman}\ \emph {et~al.}(2022)\citenamefont
  {A~Rahman}, \citenamefont {Lewis}, \citenamefont {Mendicelli},\ and\
  \citenamefont {Powell}}]{rahman2022}%
  \BibitemOpen
  \bibfield  {author} {\bibinfo {author} {\bibfnamefont {S.}~\bibnamefont
  {A~Rahman}}, \bibinfo {author} {\bibfnamefont {R.}~\bibnamefont {Lewis}},
  \bibinfo {author} {\bibfnamefont {E.}~\bibnamefont {Mendicelli}},\ and\
  \bibinfo {author} {\bibfnamefont {S.}~\bibnamefont {Powell}},\ }\bibfield
  {title} {\bibinfo {title} {Self-mitigating trotter circuits for su(2) lattice
  gauge theory on a quantum computer},\ }\href
  {https://doi.org/10.1103/PhysRevD.106.074502} {\bibfield  {journal} {\bibinfo
   {journal} {Phys. Rev. D}\ }\textbf {\bibinfo {volume} {106}},\ \bibinfo
  {pages} {074502} (\bibinfo {year} {2022})}\BibitemShut {NoStop}%
\bibitem [{\citenamefont {Fromm}\ \emph {et~al.}(2022)\citenamefont {Fromm},
  \citenamefont {Philipsen},\ and\ \citenamefont {Winterowd}}]{Fromm:2022vaj}%
  \BibitemOpen
  \bibfield  {author} {\bibinfo {author} {\bibfnamefont {M.}~\bibnamefont
  {Fromm}}, \bibinfo {author} {\bibfnamefont {O.}~\bibnamefont {Philipsen}},\
  and\ \bibinfo {author} {\bibfnamefont {C.}~\bibnamefont {Winterowd}},\
  }\bibfield  {title} {\bibinfo {title} {Dihedral lattice gauge theories on a
  quantum annealer},\ }\href@noop {} {\bibfield  {journal} {\bibinfo  {journal}
  {arXiv preprint arXiv:2206.14679}\ } (\bibinfo {year} {2022})}\BibitemShut
  {NoStop}%
\bibitem [{\citenamefont {Farrell}\ \emph
  {et~al.}(2022{\natexlab{a}})\citenamefont {Farrell}, \citenamefont
  {Chernyshev}, \citenamefont {Powell}, \citenamefont {Zemlevskiy},
  \citenamefont {Illa},\ and\ \citenamefont
  {Savage}}]{farrell_preparations_2022}%
  \BibitemOpen
  \bibfield  {author} {\bibinfo {author} {\bibfnamefont {R.~C.}\ \bibnamefont
  {Farrell}}, \bibinfo {author} {\bibfnamefont {I.~A.}\ \bibnamefont
  {Chernyshev}}, \bibinfo {author} {\bibfnamefont {S.~J.}\ \bibnamefont
  {Powell}}, \bibinfo {author} {\bibfnamefont {N.~A.}\ \bibnamefont
  {Zemlevskiy}}, \bibinfo {author} {\bibfnamefont {M.}~\bibnamefont {Illa}},\
  and\ \bibinfo {author} {\bibfnamefont {M.~J.}\ \bibnamefont {Savage}},\
  }\bibfield  {title} {\bibinfo {title} {Preparations for {{Quantum
  Simulations}} of {{Quantum Chromodynamics}} in 1+1 {{Dimensions}}: ({{I}})
  {{Axial Gauge}}},\ }\href@noop {} {\bibfield  {journal} {\bibinfo  {journal}
  {arXiv preprint arXiv:2207.01731}\ } (\bibinfo {year}
  {2022}{\natexlab{a}})}\BibitemShut {NoStop}%
\bibitem [{\citenamefont {Farrell}\ \emph
  {et~al.}(2022{\natexlab{b}})\citenamefont {Farrell}, \citenamefont
  {Chernyshev}, \citenamefont {Powell}, \citenamefont {Zemlevskiy},
  \citenamefont {Illa},\ and\ \citenamefont
  {Savage}}]{farrell2022preparations}%
  \BibitemOpen
  \bibfield  {author} {\bibinfo {author} {\bibfnamefont {R.~C.}\ \bibnamefont
  {Farrell}}, \bibinfo {author} {\bibfnamefont {I.~A.}\ \bibnamefont
  {Chernyshev}}, \bibinfo {author} {\bibfnamefont {S.~J.}\ \bibnamefont
  {Powell}}, \bibinfo {author} {\bibfnamefont {N.~A.}\ \bibnamefont
  {Zemlevskiy}}, \bibinfo {author} {\bibfnamefont {M.}~\bibnamefont {Illa}},\
  and\ \bibinfo {author} {\bibfnamefont {M.~J.}\ \bibnamefont {Savage}},\
  }\bibfield  {title} {\bibinfo {title} {Preparations for quantum simulations
  of quantum chromodynamics in 1+ 1 dimensions:(ii) single-baryon $\beta
  $-decay in real time},\ }\href@noop {} {\bibfield  {journal} {\bibinfo
  {journal} {arXiv preprint arXiv:2209.10781}\ } (\bibinfo {year}
  {2022}{\natexlab{b}})}\BibitemShut {NoStop}%
\bibitem [{\citenamefont {Kogut}\ and\ \citenamefont
  {Susskind}(1975)}]{kogut_hamiltonian_1975}%
  \BibitemOpen
  \bibfield  {author} {\bibinfo {author} {\bibfnamefont {J.}~\bibnamefont
  {Kogut}}\ and\ \bibinfo {author} {\bibfnamefont {L.}~\bibnamefont
  {Susskind}},\ }\bibfield  {title} {\bibinfo {title} {Hamiltonian formulation
  of {{Wilson}}'s lattice gauge theories},\ }\href
  {https://doi.org/10.1103/PhysRevD.11.395} {\bibfield  {journal} {\bibinfo
  {journal} {Phys. Rev. D}\ }\textbf {\bibinfo {volume} {11}},\ \bibinfo
  {pages} {395} (\bibinfo {year} {1975})}\BibitemShut {NoStop}%
\bibitem [{\citenamefont {Griffiths}(2020)}]{griffiths2020introduction}%
  \BibitemOpen
  \bibfield  {author} {\bibinfo {author} {\bibfnamefont {D.}~\bibnamefont
  {Griffiths}},\ }\href@noop {} {\emph {\bibinfo {title} {Introduction to
  elementary particles}}}\ (\bibinfo  {publisher} {John Wiley \& Sons},\
  \bibinfo {year} {2020})\BibitemShut {NoStop}%
\bibitem [{\citenamefont {Zohar}\ \emph {et~al.}(2015)\citenamefont {Zohar},
  \citenamefont {Cirac},\ and\ \citenamefont {Reznik}}]{zohar_quantum_2015}%
  \BibitemOpen
  \bibfield  {author} {\bibinfo {author} {\bibfnamefont {E.}~\bibnamefont
  {Zohar}}, \bibinfo {author} {\bibfnamefont {J.~I.}\ \bibnamefont {Cirac}},\
  and\ \bibinfo {author} {\bibfnamefont {B.}~\bibnamefont {Reznik}},\
  }\bibfield  {title} {\bibinfo {title} {Quantum simulations of lattice gauge
  theories using ultracold atoms in optical lattices},\ }\href
  {https://doi.org/10.1088/0034-4885/79/1/014401} {\bibfield  {journal}
  {\bibinfo  {journal} {Rep. Prog. Phys.}\ }\textbf {\bibinfo {volume} {79}},\
  \bibinfo {pages} {014401} (\bibinfo {year} {2015})}\BibitemShut {NoStop}%
\bibitem [{\citenamefont {K{\"u}hn}\ \emph {et~al.}(2015)\citenamefont
  {K{\"u}hn}, \citenamefont {Zohar}, \citenamefont {Cirac},\ and\ \citenamefont
  {Ba{\~n}uls}}]{kuhn_non-abelian_2015}%
  \BibitemOpen
  \bibfield  {author} {\bibinfo {author} {\bibfnamefont {S.}~\bibnamefont
  {K{\"u}hn}}, \bibinfo {author} {\bibfnamefont {E.}~\bibnamefont {Zohar}},
  \bibinfo {author} {\bibfnamefont {J.~I.}\ \bibnamefont {Cirac}},\ and\
  \bibinfo {author} {\bibfnamefont {M.~C.}\ \bibnamefont {Ba{\~n}uls}},\
  }\bibfield  {title} {\bibinfo {title} {Non-{{Abelian}} string breaking
  phenomena with matrix product states},\ }\href
  {https://doi.org/10.1007/JHEP07(2015)130} {\bibfield  {journal} {\bibinfo
  {journal} {J. High Energ. Phys.}\ }\textbf {\bibinfo {volume} {2015}}\bibinfo
   {number} { (7)},\ \bibinfo {pages} {130}}\BibitemShut {NoStop}%
\bibitem [{\citenamefont {{Jordan}}\ and\ \citenamefont
  {{Wigner}}(1928)}]{JordanWigner}%
  \BibitemOpen
\bibfield  {number} {  }\bibfield  {author} {\bibinfo {author} {\bibfnamefont
  {P.}~\bibnamefont {{Jordan}}}\ and\ \bibinfo {author} {\bibfnamefont
  {E.}~\bibnamefont {{Wigner}}},\ }\bibfield  {title} {\bibinfo {title}
  {{{\"U}ber das Paulische {\"A}quivalenzverbot}},\ }\href
  {https://doi.org/10.1007/BF01331938} {\bibfield  {journal} {\bibinfo
  {journal} {Z. Physik}\ }\textbf {\bibinfo {volume} {47}},\ \bibinfo {pages}
  {631} (\bibinfo {year} {1928})}\BibitemShut {NoStop}%
\bibitem [{\citenamefont {Efron}(1979)}]{bootstrap}%
  \BibitemOpen
  \bibfield  {author} {\bibinfo {author} {\bibfnamefont {B.}~\bibnamefont
  {Efron}},\ }\bibfield  {title} {\bibinfo {title} {{Bootstrap Methods: Another
  Look at the Jackknife}},\ }\href {https://doi.org/10.1214/aos/1176344552}
  {\bibfield  {journal} {\bibinfo  {journal} {Ann. Stat.}\ }\textbf {\bibinfo
  {volume} {7}},\ \bibinfo {pages} {1 } (\bibinfo {year} {1979})}\BibitemShut
  {NoStop}%
\bibitem [{\citenamefont {Lloyd}(1996)}]{lloyd1996}%
  \BibitemOpen
  \bibfield  {author} {\bibinfo {author} {\bibfnamefont {S.}~\bibnamefont
  {Lloyd}},\ }\bibfield  {title} {\bibinfo {title} {Universal quantum
  simulators},\ }\href@noop {} {\bibfield  {journal} {\bibinfo  {journal}
  {Science}\ }\textbf {\bibinfo {volume} {273}},\ \bibinfo {pages} {1073}
  (\bibinfo {year} {1996})}\BibitemShut {NoStop}%
\bibitem [{\citenamefont {Hashim}\ \emph {et~al.}(2021)\citenamefont {Hashim},
  \citenamefont {Naik}, \citenamefont {Morvan}, \citenamefont {Ville},
  \citenamefont {Mitchell}, \citenamefont {Kreikebaum}, \citenamefont {Davis},
  \citenamefont {Smith}, \citenamefont {Iancu}, \citenamefont {O'Brien},
  \citenamefont {Hincks}, \citenamefont {Wallman}, \citenamefont {Emerson},\
  and\ \citenamefont {Siddiqi}}]{Hashim2021Randomized}%
  \BibitemOpen
  \bibfield  {author} {\bibinfo {author} {\bibfnamefont {A.}~\bibnamefont
  {Hashim}}, \bibinfo {author} {\bibfnamefont {R.~K.}\ \bibnamefont {Naik}},
  \bibinfo {author} {\bibfnamefont {A.}~\bibnamefont {Morvan}}, \bibinfo
  {author} {\bibfnamefont {J.-L.}\ \bibnamefont {Ville}}, \bibinfo {author}
  {\bibfnamefont {B.}~\bibnamefont {Mitchell}}, \bibinfo {author}
  {\bibfnamefont {J.~M.}\ \bibnamefont {Kreikebaum}}, \bibinfo {author}
  {\bibfnamefont {M.}~\bibnamefont {Davis}}, \bibinfo {author} {\bibfnamefont
  {E.}~\bibnamefont {Smith}}, \bibinfo {author} {\bibfnamefont
  {C.}~\bibnamefont {Iancu}}, \bibinfo {author} {\bibfnamefont {K.~P.}\
  \bibnamefont {O'Brien}}, \bibinfo {author} {\bibfnamefont {I.}~\bibnamefont
  {Hincks}}, \bibinfo {author} {\bibfnamefont {J.~J.}\ \bibnamefont {Wallman}},
  \bibinfo {author} {\bibfnamefont {J.}~\bibnamefont {Emerson}},\ and\ \bibinfo
  {author} {\bibfnamefont {I.}~\bibnamefont {Siddiqi}},\ }\bibfield  {title}
  {\bibinfo {title} {Randomized compiling for scalable quantum computing on a
  noisy superconducting quantum processor},\ }\href@noop {} {\bibfield
  {journal} {\bibinfo  {journal} {Phys. Rev. X}\ }\textbf {\bibinfo {volume}
  {11}},\ \bibinfo {pages} {041039} (\bibinfo {year} {2021})}\BibitemShut
  {NoStop}%
\bibitem [{IBM(2022)}]{IBMQ}%
  \BibitemOpen
  \href@noop {} {}\bibinfo {howpublished} {IBM Quantum.
  \url{https://quantum-computing.ibm.com/}} (\bibinfo {year}
  {2022})\BibitemShut {NoStop}%
\bibitem [{\citenamefont {Kruschke}(2014)}]{kruschke2014}%
  \BibitemOpen
  \bibfield  {author} {\bibinfo {author} {\bibfnamefont {J.}~\bibnamefont
  {Kruschke}},\ }\href@noop {} {\emph {\bibinfo {title} {Doing Bayesian Data
  Analysis: A Tutorial with R, JAGS, and Stan}}},\ \bibinfo {edition} {2nd}\
  ed.\ (\bibinfo  {publisher} {Academic Press},\ \bibinfo {year}
  {2014})\BibitemShut {NoStop}%
\bibitem [{\citenamefont {Hamer}\ \emph {et~al.}(1997)\citenamefont {Hamer},
  \citenamefont {Weihong},\ and\ \citenamefont {Oitmaa}}]{hamer_series_1997}%
  \BibitemOpen
  \bibfield  {author} {\bibinfo {author} {\bibfnamefont {C.~J.}\ \bibnamefont
  {Hamer}}, \bibinfo {author} {\bibfnamefont {Z.}~\bibnamefont {Weihong}},\
  and\ \bibinfo {author} {\bibfnamefont {J.}~\bibnamefont {Oitmaa}},\
  }\bibfield  {title} {\bibinfo {title} {Series expansions for the massive
  {{Schwinger}} model in {{Hamiltonian}} lattice theory},\ }\href
  {https://doi.org/10.1103/PhysRevD.56.55} {\bibfield  {journal} {\bibinfo
  {journal} {Phys. Rev. D}\ }\textbf {\bibinfo {volume} {56}},\ \bibinfo
  {pages} {55} (\bibinfo {year} {1997})}\BibitemShut {NoStop}%
\bibitem [{\citenamefont {Muschik}\ \emph {et~al.}(2017)\citenamefont
  {Muschik}, \citenamefont {Heyl}, \citenamefont {Martinez}, \citenamefont
  {Monz}, \citenamefont {Schindler}, \citenamefont {Vogell}, \citenamefont
  {{Marcello Dalmonte}}, \citenamefont {Hauke}, \citenamefont {Blatt},\ and\
  \citenamefont {Zoller}}]{muschik_u1_2017}%
  \BibitemOpen
  \bibfield  {author} {\bibinfo {author} {\bibfnamefont {C.}~\bibnamefont
  {Muschik}}, \bibinfo {author} {\bibfnamefont {M.}~\bibnamefont {Heyl}},
  \bibinfo {author} {\bibfnamefont {E.}~\bibnamefont {Martinez}}, \bibinfo
  {author} {\bibfnamefont {T.}~\bibnamefont {Monz}}, \bibinfo {author}
  {\bibfnamefont {P.}~\bibnamefont {Schindler}}, \bibinfo {author}
  {\bibfnamefont {B.}~\bibnamefont {Vogell}}, \bibinfo {author} {\bibnamefont
  {{Marcello Dalmonte}}}, \bibinfo {author} {\bibfnamefont {P.}~\bibnamefont
  {Hauke}}, \bibinfo {author} {\bibfnamefont {R.}~\bibnamefont {Blatt}},\ and\
  \bibinfo {author} {\bibfnamefont {P.}~\bibnamefont {Zoller}},\ }\bibfield
  {title} {\bibinfo {title} {U(1) {{Wilson}} lattice gauge theories in digital
  quantum simulators},\ }\href {https://doi.org/10.1088/1367-2630/aa89ab}
  {\bibfield  {journal} {\bibinfo  {journal} {New J. Phys.}\ }\textbf {\bibinfo
  {volume} {19}},\ \bibinfo {pages} {103020} (\bibinfo {year}
  {2017})}\BibitemShut {NoStop}%
\bibitem [{\citenamefont {Jordan}\ \emph {et~al.}(2012)\citenamefont {Jordan},
  \citenamefont {Lee},\ and\ \citenamefont {Preskill}}]{jordan_quantum_2012}%
  \BibitemOpen
  \bibfield  {author} {\bibinfo {author} {\bibfnamefont {S.~P.}\ \bibnamefont
  {Jordan}}, \bibinfo {author} {\bibfnamefont {K.~S.~M.}\ \bibnamefont {Lee}},\
  and\ \bibinfo {author} {\bibfnamefont {J.}~\bibnamefont {Preskill}},\
  }\bibfield  {title} {\bibinfo {title} {Quantum {{Algorithms}} for {{Quantum
  Field Theories}}},\ }\href {https://doi.org/10.1126/science.1217069}
  {\bibfield  {journal} {\bibinfo  {journal} {Science}\ }\textbf {\bibinfo
  {volume} {336}},\ \bibinfo {pages} {1130} (\bibinfo {year}
  {2012})}\BibitemShut {NoStop}%
\bibitem [{\citenamefont {Jordan}\ \emph {et~al.}(2014)\citenamefont {Jordan},
  \citenamefont {Lee},\ and\ \citenamefont {Preskill}}]{jordan2014quantum}%
  \BibitemOpen
  \bibfield  {author} {\bibinfo {author} {\bibfnamefont {S.~P.}\ \bibnamefont
  {Jordan}}, \bibinfo {author} {\bibfnamefont {K.~S.}\ \bibnamefont {Lee}},\
  and\ \bibinfo {author} {\bibfnamefont {J.}~\bibnamefont {Preskill}},\
  }\bibfield  {title} {\bibinfo {title} {Quantum algorithms for fermionic
  quantum field theories},\ }\href@noop {} {\bibfield  {journal} {\bibinfo
  {journal} {arXiv preprint arXiv:1404.7115}\ } (\bibinfo {year}
  {2014})}\BibitemShut {NoStop}%
\bibitem [{\citenamefont {Tong}\ \emph {et~al.}(2021)\citenamefont {Tong},
  \citenamefont {Albert}, \citenamefont {McClean}, \citenamefont {Preskill},\
  and\ \citenamefont {Su}}]{tong2021provably}%
  \BibitemOpen
  \bibfield  {author} {\bibinfo {author} {\bibfnamefont {Y.}~\bibnamefont
  {Tong}}, \bibinfo {author} {\bibfnamefont {V.~V.}\ \bibnamefont {Albert}},
  \bibinfo {author} {\bibfnamefont {J.~R.}\ \bibnamefont {McClean}}, \bibinfo
  {author} {\bibfnamefont {J.}~\bibnamefont {Preskill}},\ and\ \bibinfo
  {author} {\bibfnamefont {Y.}~\bibnamefont {Su}},\ }\bibfield  {title}
  {\bibinfo {title} {Provably accurate simulation of gauge theories and bosonic
  systems},\ }\href@noop {} {\bibfield  {journal} {\bibinfo  {journal} {arXiv
  preprint arXiv:2110.06942}\ } (\bibinfo {year} {2021})}\BibitemShut {NoStop}%
\bibitem [{\citenamefont {Ott}\ \emph {et~al.}(2021)\citenamefont {Ott},
  \citenamefont {Zache}, \citenamefont {Jendrzejewski},\ and\ \citenamefont
  {Berges}}]{ott2021scalable}%
  \BibitemOpen
  \bibfield  {author} {\bibinfo {author} {\bibfnamefont {R.}~\bibnamefont
  {Ott}}, \bibinfo {author} {\bibfnamefont {T.~V.}\ \bibnamefont {Zache}},
  \bibinfo {author} {\bibfnamefont {F.}~\bibnamefont {Jendrzejewski}},\ and\
  \bibinfo {author} {\bibfnamefont {J.}~\bibnamefont {Berges}},\ }\bibfield
  {title} {\bibinfo {title} {Scalable cold-atom quantum simulator for
  two-dimensional {{QED}}},\ }\href@noop {} {\bibfield  {journal} {\bibinfo
  {journal} {Physical Review Letters}\ }\textbf {\bibinfo {volume} {127}},\
  \bibinfo {pages} {130504} (\bibinfo {year} {2021})}\BibitemShut {NoStop}%
\bibitem [{\citenamefont {Kasper}\ \emph {et~al.}(2017)\citenamefont {Kasper},
  \citenamefont {Hebenstreit}, \citenamefont {Jendrzejewski}, \citenamefont
  {Oberthaler},\ and\ \citenamefont {Berges}}]{kasper2017implementing}%
  \BibitemOpen
  \bibfield  {author} {\bibinfo {author} {\bibfnamefont {V.}~\bibnamefont
  {Kasper}}, \bibinfo {author} {\bibfnamefont {F.}~\bibnamefont {Hebenstreit}},
  \bibinfo {author} {\bibfnamefont {F.}~\bibnamefont {Jendrzejewski}}, \bibinfo
  {author} {\bibfnamefont {M.~K.}\ \bibnamefont {Oberthaler}},\ and\ \bibinfo
  {author} {\bibfnamefont {J.}~\bibnamefont {Berges}},\ }\bibfield  {title}
  {\bibinfo {title} {Implementing quantum electrodynamics with ultracold atomic
  systems},\ }\href@noop {} {\bibfield  {journal} {\bibinfo  {journal} {New
  journal of physics}\ }\textbf {\bibinfo {volume} {19}},\ \bibinfo {pages}
  {023030} (\bibinfo {year} {2017})}\BibitemShut {NoStop}%
\bibitem [{\citenamefont {Kasper}\ \emph {et~al.}(2016)\citenamefont {Kasper},
  \citenamefont {Hebenstreit}, \citenamefont {Oberthaler},\ and\ \citenamefont
  {Berges}}]{kasper2016schwinger}%
  \BibitemOpen
  \bibfield  {author} {\bibinfo {author} {\bibfnamefont {V.}~\bibnamefont
  {Kasper}}, \bibinfo {author} {\bibfnamefont {F.}~\bibnamefont {Hebenstreit}},
  \bibinfo {author} {\bibfnamefont {M.~K.}\ \bibnamefont {Oberthaler}},\ and\
  \bibinfo {author} {\bibfnamefont {J.}~\bibnamefont {Berges}},\ }\bibfield
  {title} {\bibinfo {title} {Schwinger pair production with ultracold atoms},\
  }\href@noop {} {\bibfield  {journal} {\bibinfo  {journal} {Physics Letters
  B}\ }\textbf {\bibinfo {volume} {760}},\ \bibinfo {pages} {742} (\bibinfo
  {year} {2016})}\BibitemShut {NoStop}%
\bibitem [{\citenamefont {Andrade}\ \emph {et~al.}(2022)\citenamefont
  {Andrade}, \citenamefont {Davoudi}, \citenamefont {Gra{\ss}}, \citenamefont
  {Hafezi}, \citenamefont {Pagano},\ and\ \citenamefont
  {Seif}}]{andrade2022engineering}%
  \BibitemOpen
  \bibfield  {author} {\bibinfo {author} {\bibfnamefont {B.}~\bibnamefont
  {Andrade}}, \bibinfo {author} {\bibfnamefont {Z.}~\bibnamefont {Davoudi}},
  \bibinfo {author} {\bibfnamefont {T.}~\bibnamefont {Gra{\ss}}}, \bibinfo
  {author} {\bibfnamefont {M.}~\bibnamefont {Hafezi}}, \bibinfo {author}
  {\bibfnamefont {G.}~\bibnamefont {Pagano}},\ and\ \bibinfo {author}
  {\bibfnamefont {A.}~\bibnamefont {Seif}},\ }\bibfield  {title} {\bibinfo
  {title} {Engineering an effective three-spin hamiltonian in trapped-ion
  systems for applications in quantum simulation},\ }\href@noop {} {\bibfield
  {journal} {\bibinfo  {journal} {Quantum Science and Technology}\ }\textbf
  {\bibinfo {volume} {7}},\ \bibinfo {pages} {034001} (\bibinfo {year}
  {2022})}\BibitemShut {NoStop}%
\bibitem [{\citenamefont {Davoudi}\ \emph
  {et~al.}(2021{\natexlab{a}})\citenamefont {Davoudi}, \citenamefont {Linke},\
  and\ \citenamefont {Pagano}}]{davoudi2021toward}%
  \BibitemOpen
  \bibfield  {author} {\bibinfo {author} {\bibfnamefont {Z.}~\bibnamefont
  {Davoudi}}, \bibinfo {author} {\bibfnamefont {N.~M.}\ \bibnamefont {Linke}},\
  and\ \bibinfo {author} {\bibfnamefont {G.}~\bibnamefont {Pagano}},\
  }\bibfield  {title} {\bibinfo {title} {Toward simulating quantum field
  theories with controlled phonon-ion dynamics: A hybrid analog-digital
  approach},\ }\href@noop {} {\bibfield  {journal} {\bibinfo  {journal}
  {Physical Review Research}\ }\textbf {\bibinfo {volume} {3}},\ \bibinfo
  {pages} {043072} (\bibinfo {year} {2021}{\natexlab{a}})}\BibitemShut
  {NoStop}%
\bibitem [{\citenamefont {Tagliacozzo}\ \emph {et~al.}(2013)\citenamefont
  {Tagliacozzo}, \citenamefont {Celi}, \citenamefont {Orland}, \citenamefont
  {Mitchell},\ and\ \citenamefont {Lewenstein}}]{tagliacozzo_simulation_2013}%
  \BibitemOpen
  \bibfield  {author} {\bibinfo {author} {\bibfnamefont {L.}~\bibnamefont
  {Tagliacozzo}}, \bibinfo {author} {\bibfnamefont {A.}~\bibnamefont {Celi}},
  \bibinfo {author} {\bibfnamefont {P.}~\bibnamefont {Orland}}, \bibinfo
  {author} {\bibfnamefont {M.}~\bibnamefont {Mitchell}},\ and\ \bibinfo
  {author} {\bibfnamefont {M.}~\bibnamefont {Lewenstein}},\ }\bibfield  {title}
  {\bibinfo {title} {{Simulation of non-Abelian gauge theories with optical
  lattices}},\ }\href@noop {} {\bibfield  {journal} {\bibinfo  {journal} {Nat.
  Commun.}\ }\textbf {\bibinfo {volume} {4}},\ \bibinfo {pages} {1} (\bibinfo
  {year} {2013})}\BibitemShut {NoStop}%
\bibitem [{\citenamefont {Mezzacapo}\ \emph {et~al.}(2015)\citenamefont
  {Mezzacapo}, \citenamefont {Rico}, \citenamefont {Sab{\'i}n}, \citenamefont
  {Egusquiza}, \citenamefont {Lamata},\ and\ \citenamefont
  {Solano}}]{mezzacapo_non-abelian_2015}%
  \BibitemOpen
  \bibfield  {author} {\bibinfo {author} {\bibfnamefont {A.}~\bibnamefont
  {Mezzacapo}}, \bibinfo {author} {\bibfnamefont {E.}~\bibnamefont {Rico}},
  \bibinfo {author} {\bibfnamefont {C.}~\bibnamefont {Sab{\'i}n}}, \bibinfo
  {author} {\bibfnamefont {I.~L.}\ \bibnamefont {Egusquiza}}, \bibinfo {author}
  {\bibfnamefont {L.}~\bibnamefont {Lamata}},\ and\ \bibinfo {author}
  {\bibfnamefont {E.}~\bibnamefont {Solano}},\ }\bibfield  {title} {\bibinfo
  {title} {Non-{{Abelian SU}}(2) {{Lattice Gauge Theories}} in
  {{Superconducting Circuits}}},\ }\href
  {https://doi.org/10.1103/PhysRevLett.115.240502} {\bibfield  {journal}
  {\bibinfo  {journal} {Phys. Rev. Lett.}\ }\textbf {\bibinfo {volume} {115}},\
  \bibinfo {pages} {240502} (\bibinfo {year} {2015})}\BibitemShut {NoStop}%
\bibitem [{\citenamefont {Bender}\ \emph {et~al.}(2018)\citenamefont {Bender},
  \citenamefont {Zohar}, \citenamefont {Farace},\ and\ \citenamefont
  {Cirac}}]{bender_digital_2018}%
  \BibitemOpen
  \bibfield  {author} {\bibinfo {author} {\bibfnamefont {J.}~\bibnamefont
  {Bender}}, \bibinfo {author} {\bibfnamefont {E.}~\bibnamefont {Zohar}},
  \bibinfo {author} {\bibfnamefont {A.}~\bibnamefont {Farace}},\ and\ \bibinfo
  {author} {\bibfnamefont {J.~I.}\ \bibnamefont {Cirac}},\ }\bibfield  {title}
  {\bibinfo {title} {Digital quantum simulation of lattice gauge theories in
  three spatial dimensions},\ }\href {https://doi.org/10.1088/1367-2630/aadb71}
  {\bibfield  {journal} {\bibinfo  {journal} {New J. Phys.}\ }\textbf {\bibinfo
  {volume} {20}},\ \bibinfo {pages} {093001} (\bibinfo {year}
  {2018})}\BibitemShut {NoStop}%
\bibitem [{\citenamefont {Zohar}\ and\ \citenamefont
  {Cirac}(2018)}]{zohar_eliminating_2018}%
  \BibitemOpen
  \bibfield  {author} {\bibinfo {author} {\bibfnamefont {E.}~\bibnamefont
  {Zohar}}\ and\ \bibinfo {author} {\bibfnamefont {J.~I.}\ \bibnamefont
  {Cirac}},\ }\bibfield  {title} {\bibinfo {title} {Eliminating fermionic
  matter fields in lattice gauge theories},\ }\href
  {https://doi.org/10.1103/PhysRevB.98.075119} {\bibfield  {journal} {\bibinfo
  {journal} {Phys. Rev. B}\ }\textbf {\bibinfo {volume} {98}},\ \bibinfo
  {pages} {075119} (\bibinfo {year} {2018})}\BibitemShut {NoStop}%
\bibitem [{\citenamefont {Zache}\ \emph {et~al.}(2019)\citenamefont {Zache},
  \citenamefont {Mueller}, \citenamefont {Schneider}, \citenamefont
  {Jendrzejewski}, \citenamefont {Berges},\ and\ \citenamefont
  {Hauke}}]{zache_dynamical_2019}%
  \BibitemOpen
  \bibfield  {author} {\bibinfo {author} {\bibfnamefont {T.}~\bibnamefont
  {Zache}}, \bibinfo {author} {\bibfnamefont {N.}~\bibnamefont {Mueller}},
  \bibinfo {author} {\bibfnamefont {J.}~\bibnamefont {Schneider}}, \bibinfo
  {author} {\bibfnamefont {F.}~\bibnamefont {Jendrzejewski}}, \bibinfo {author}
  {\bibfnamefont {J.}~\bibnamefont {Berges}},\ and\ \bibinfo {author}
  {\bibfnamefont {P.}~\bibnamefont {Hauke}},\ }\bibfield  {title} {\bibinfo
  {title} {Dynamical topological transitions in the massive {{Schwinger}} model
  with a $\theta$ term},\ }\href@noop {} {\bibfield  {journal} {\bibinfo
  {journal} {Physical review letters}\ }\textbf {\bibinfo {volume} {122}},\
  \bibinfo {pages} {050403} (\bibinfo {year} {2019})}\BibitemShut {NoStop}%
\bibitem [{\citenamefont {Kasper}\ \emph
  {et~al.}(2020{\natexlab{a}})\citenamefont {Kasper}, \citenamefont {Zache},
  \citenamefont {Jendrzejewski}, \citenamefont {Lewenstein},\ and\
  \citenamefont {Zohar}}]{kasper_non-abelian_2020}%
  \BibitemOpen
  \bibfield  {author} {\bibinfo {author} {\bibfnamefont {V.}~\bibnamefont
  {Kasper}}, \bibinfo {author} {\bibfnamefont {T.~V.}\ \bibnamefont {Zache}},
  \bibinfo {author} {\bibfnamefont {F.}~\bibnamefont {Jendrzejewski}}, \bibinfo
  {author} {\bibfnamefont {M.}~\bibnamefont {Lewenstein}},\ and\ \bibinfo
  {author} {\bibfnamefont {E.}~\bibnamefont {Zohar}},\ }\bibfield  {title}
  {\bibinfo {title} {Non--{{Abelian}} gauge invariance from dynamical
  decoupling},\ }\href@noop {} {\bibfield  {journal} {\bibinfo  {journal}
  {arXiv preprint arXiv:2012.08620}\ } (\bibinfo {year}
  {2020}{\natexlab{a}})}\BibitemShut {NoStop}%
\bibitem [{\citenamefont {Kasper}\ \emph
  {et~al.}(2020{\natexlab{b}})\citenamefont {Kasper}, \citenamefont
  {Juzeli{\=u}nas}, \citenamefont {Lewenstein}, \citenamefont {Jendrzejewski},\
  and\ \citenamefont {Zohar}}]{kasper_jaynescummings_2020}%
  \BibitemOpen
  \bibfield  {author} {\bibinfo {author} {\bibfnamefont {V.}~\bibnamefont
  {Kasper}}, \bibinfo {author} {\bibfnamefont {G.}~\bibnamefont
  {Juzeli{\=u}nas}}, \bibinfo {author} {\bibfnamefont {M.}~\bibnamefont
  {Lewenstein}}, \bibinfo {author} {\bibfnamefont {F.}~\bibnamefont
  {Jendrzejewski}},\ and\ \bibinfo {author} {\bibfnamefont {E.}~\bibnamefont
  {Zohar}},\ }\bibfield  {title} {\bibinfo {title} {From the
  {{Jaynes}}\textendash{{Cummings}} model to non-abelian gauge theories: A
  guided tour for the quantum engineer},\ }\href
  {https://doi.org/10.1088/1367-2630/abb961} {\bibfield  {journal} {\bibinfo
  {journal} {New J. Phys.}\ }\textbf {\bibinfo {volume} {22}},\ \bibinfo
  {pages} {103027} (\bibinfo {year} {2020}{\natexlab{b}})}\BibitemShut
  {NoStop}%
\bibitem [{\citenamefont {Raychowdhury}\ and\ \citenamefont
  {Stryker}(2020)}]{raychowdhury_loop_2020}%
  \BibitemOpen
  \bibfield  {author} {\bibinfo {author} {\bibfnamefont {I.}~\bibnamefont
  {Raychowdhury}}\ and\ \bibinfo {author} {\bibfnamefont {J.~R.}\ \bibnamefont
  {Stryker}},\ }\bibfield  {title} {\bibinfo {title} {Loop, string, and hadron
  dynamics in {{SU}}(2) {{Hamiltonian}} lattice gauge theories},\ }\href
  {https://doi.org/10.1103/PhysRevD.101.114502} {\bibfield  {journal} {\bibinfo
   {journal} {Phys. Rev. D}\ }\textbf {\bibinfo {volume} {101}},\ \bibinfo
  {pages} {114502} (\bibinfo {year} {2020})}\BibitemShut {NoStop}%
\bibitem [{\citenamefont {Davoudi}\ \emph
  {et~al.}(2021{\natexlab{b}})\citenamefont {Davoudi}, \citenamefont
  {Raychowdhury},\ and\ \citenamefont {Shaw}}]{davoudi_search_2021}%
  \BibitemOpen
  \bibfield  {author} {\bibinfo {author} {\bibfnamefont {Z.}~\bibnamefont
  {Davoudi}}, \bibinfo {author} {\bibfnamefont {I.}~\bibnamefont
  {Raychowdhury}},\ and\ \bibinfo {author} {\bibfnamefont {A.}~\bibnamefont
  {Shaw}},\ }\bibfield  {title} {\bibinfo {title} {Search for efficient
  formulations for {{Hamiltonian}} simulation of non-{{Abelian}} lattice gauge
  theories},\ }\href {https://doi.org/10.1103/PhysRevD.104.074505} {\bibfield
  {journal} {\bibinfo  {journal} {Phys. Rev. D}\ }\textbf {\bibinfo {volume}
  {104}},\ \bibinfo {pages} {074505} (\bibinfo {year}
  {2021}{\natexlab{b}})}\BibitemShut {NoStop}%
\bibitem [{\citenamefont {Davoudi}\ \emph {et~al.}(2020)\citenamefont
  {Davoudi}, \citenamefont {Hafezi}, \citenamefont {Monroe}, \citenamefont
  {Pagano}, \citenamefont {Seif},\ and\ \citenamefont
  {Shaw}}]{davoudi2020towards}%
  \BibitemOpen
  \bibfield  {author} {\bibinfo {author} {\bibfnamefont {Z.}~\bibnamefont
  {Davoudi}}, \bibinfo {author} {\bibfnamefont {M.}~\bibnamefont {Hafezi}},
  \bibinfo {author} {\bibfnamefont {C.}~\bibnamefont {Monroe}}, \bibinfo
  {author} {\bibfnamefont {G.}~\bibnamefont {Pagano}}, \bibinfo {author}
  {\bibfnamefont {A.}~\bibnamefont {Seif}},\ and\ \bibinfo {author}
  {\bibfnamefont {A.}~\bibnamefont {Shaw}},\ }\bibfield  {title} {\bibinfo
  {title} {Towards analog quantum simulations of lattice gauge theories with
  trapped ions},\ }\href@noop {} {\bibfield  {journal} {\bibinfo  {journal}
  {Physical Review Research}\ }\textbf {\bibinfo {volume} {2}},\ \bibinfo
  {pages} {023015} (\bibinfo {year} {2020})}\BibitemShut {NoStop}%
\bibitem [{\citenamefont {Rico}\ \emph {et~al.}(2018)\citenamefont {Rico},
  \citenamefont {Dalmonte}, \citenamefont {Zoller}, \citenamefont {Banerjee},
  \citenamefont {Bogli}, \citenamefont {Stebler},\ and\ \citenamefont
  {Wiese}}]{rico_so3_2018}%
  \BibitemOpen
  \bibfield  {author} {\bibinfo {author} {\bibfnamefont {E.}~\bibnamefont
  {Rico}}, \bibinfo {author} {\bibfnamefont {M.}~\bibnamefont {Dalmonte}},
  \bibinfo {author} {\bibfnamefont {P.}~\bibnamefont {Zoller}}, \bibinfo
  {author} {\bibfnamefont {D.}~\bibnamefont {Banerjee}}, \bibinfo {author}
  {\bibfnamefont {M.}~\bibnamefont {Bogli}}, \bibinfo {author} {\bibfnamefont
  {P.}~\bibnamefont {Stebler}},\ and\ \bibinfo {author} {\bibfnamefont {U.-J.}\
  \bibnamefont {Wiese}},\ }\bibfield  {title} {\bibinfo {title} {{{SO}}(3)
  "{{Nuclear Physics}}" with ultracold {{Gases}}},\ }\href
  {https://doi.org/10.1016/j.aop.2018.03.020} {\bibfield  {journal} {\bibinfo
  {journal} {Annals of Physics}\ }\textbf {\bibinfo {volume} {393}},\ \bibinfo
  {pages} {466} (\bibinfo {year} {2018})}\BibitemShut {NoStop}%
\bibitem [{\citenamefont {Laflamme}\ \emph {et~al.}(2016)\citenamefont
  {Laflamme}, \citenamefont {Evans}, \citenamefont {Dalmonte}, \citenamefont
  {Gerber}, \citenamefont {Mej{\'\i}a-D{\'\i}az}, \citenamefont {Bietenholz},
  \citenamefont {Wiese},\ and\ \citenamefont {Zoller}}]{laflamme2016cp}%
  \BibitemOpen
  \bibfield  {author} {\bibinfo {author} {\bibfnamefont {C.}~\bibnamefont
  {Laflamme}}, \bibinfo {author} {\bibfnamefont {W.}~\bibnamefont {Evans}},
  \bibinfo {author} {\bibfnamefont {M.}~\bibnamefont {Dalmonte}}, \bibinfo
  {author} {\bibfnamefont {U.}~\bibnamefont {Gerber}}, \bibinfo {author}
  {\bibfnamefont {H.}~\bibnamefont {Mej{\'\i}a-D{\'\i}az}}, \bibinfo {author}
  {\bibfnamefont {W.}~\bibnamefont {Bietenholz}}, \bibinfo {author}
  {\bibfnamefont {U.-J.}\ \bibnamefont {Wiese}},\ and\ \bibinfo {author}
  {\bibfnamefont {P.}~\bibnamefont {Zoller}},\ }\bibfield  {title} {\bibinfo
  {title} {{{CP (N- 1)}} quantum field theories with alkaline-earth atoms in
  optical lattices},\ }\href@noop {} {\bibfield  {journal} {\bibinfo  {journal}
  {Annals of physics}\ }\textbf {\bibinfo {volume} {370}},\ \bibinfo {pages}
  {117} (\bibinfo {year} {2016})}\BibitemShut {NoStop}%
\bibitem [{\citenamefont {Banerjee}\ \emph {et~al.}(2022)\citenamefont
  {Banerjee}, \citenamefont {Caspar}, \citenamefont {Jiang}, \citenamefont
  {Peng},\ and\ \citenamefont {Wiese}}]{banerjee2022nematic}%
  \BibitemOpen
  \bibfield  {author} {\bibinfo {author} {\bibfnamefont {D.}~\bibnamefont
  {Banerjee}}, \bibinfo {author} {\bibfnamefont {S.}~\bibnamefont {Caspar}},
  \bibinfo {author} {\bibfnamefont {F.-J.}\ \bibnamefont {Jiang}}, \bibinfo
  {author} {\bibfnamefont {J.-H.}\ \bibnamefont {Peng}},\ and\ \bibinfo
  {author} {\bibfnamefont {U.-J.}\ \bibnamefont {Wiese}},\ }\bibfield  {title}
  {\bibinfo {title} {Nematic confined phases in the {{U(1)}} quantum link model
  on a triangular lattice: Near-term quantum computations of string dynamics on
  a chip},\ }\href@noop {} {\bibfield  {journal} {\bibinfo  {journal} {Physical
  Review Research}\ }\textbf {\bibinfo {volume} {4}},\ \bibinfo {pages}
  {023176} (\bibinfo {year} {2022})}\BibitemShut {NoStop}%
\bibitem [{\citenamefont {Marcos}\ \emph {et~al.}(2014)\citenamefont {Marcos},
  \citenamefont {Widmer}, \citenamefont {Rico}, \citenamefont {Hafezi},
  \citenamefont {Rabl}, \citenamefont {Wiese},\ and\ \citenamefont
  {Zoller}}]{marcos2014two}%
  \BibitemOpen
  \bibfield  {author} {\bibinfo {author} {\bibfnamefont {D.}~\bibnamefont
  {Marcos}}, \bibinfo {author} {\bibfnamefont {P.}~\bibnamefont {Widmer}},
  \bibinfo {author} {\bibfnamefont {E.}~\bibnamefont {Rico}}, \bibinfo {author}
  {\bibfnamefont {M.}~\bibnamefont {Hafezi}}, \bibinfo {author} {\bibfnamefont
  {P.}~\bibnamefont {Rabl}}, \bibinfo {author} {\bibfnamefont {U.-J.}\
  \bibnamefont {Wiese}},\ and\ \bibinfo {author} {\bibfnamefont
  {P.}~\bibnamefont {Zoller}},\ }\bibfield  {title} {\bibinfo {title}
  {Two-dimensional lattice gauge theories with superconducting quantum
  circuits},\ }\href@noop {} {\bibfield  {journal} {\bibinfo  {journal} {Annals
  of physics}\ }\textbf {\bibinfo {volume} {351}},\ \bibinfo {pages} {634}
  (\bibinfo {year} {2014})}\BibitemShut {NoStop}%
\bibitem [{\citenamefont {Kruckenhauser}\ \emph {et~al.}(2022)\citenamefont
  {Kruckenhauser}, \citenamefont {van Bijnen}, \citenamefont {Zache},
  \citenamefont {Di~Liberto},\ and\ \citenamefont
  {Zoller}}]{kruckenhauser2022high}%
  \BibitemOpen
  \bibfield  {author} {\bibinfo {author} {\bibfnamefont {A.}~\bibnamefont
  {Kruckenhauser}}, \bibinfo {author} {\bibfnamefont {R.}~\bibnamefont {van
  Bijnen}}, \bibinfo {author} {\bibfnamefont {T.~V.}\ \bibnamefont {Zache}},
  \bibinfo {author} {\bibfnamefont {M.}~\bibnamefont {Di~Liberto}},\ and\
  \bibinfo {author} {\bibfnamefont {P.}~\bibnamefont {Zoller}},\ }\bibfield
  {title} {\bibinfo {title} {High-dimensional {{SO(4)}}-symmetric rydberg
  manifolds for quantum simulation},\ }\href@noop {} {\bibfield  {journal}
  {\bibinfo  {journal} {arXiv preprint arXiv:2206.01108}\ } (\bibinfo {year}
  {2022})}\BibitemShut {NoStop}%
\bibitem [{\citenamefont {Gonz{\'a}lez-Cuadra}\ \emph
  {et~al.}(2022)\citenamefont {Gonz{\'a}lez-Cuadra}, \citenamefont {Zache},
  \citenamefont {Carrasco}, \citenamefont {Kraus},\ and\ \citenamefont
  {Zoller}}]{gonzalez-cuadra_hardware_2022}%
  \BibitemOpen
  \bibfield  {author} {\bibinfo {author} {\bibfnamefont {D.}~\bibnamefont
  {Gonz{\'a}lez-Cuadra}}, \bibinfo {author} {\bibfnamefont {T.~V.}\
  \bibnamefont {Zache}}, \bibinfo {author} {\bibfnamefont {J.}~\bibnamefont
  {Carrasco}}, \bibinfo {author} {\bibfnamefont {B.}~\bibnamefont {Kraus}},\
  and\ \bibinfo {author} {\bibfnamefont {P.}~\bibnamefont {Zoller}},\
  }\bibfield  {title} {\bibinfo {title} {Hardware efficient quantum simulation
  of non-abelian gauge theories with qudits on {{Rydberg}} platforms},\
  }\href@noop {} {\bibfield  {journal} {\bibinfo  {journal} {arXiv preprint
  arXiv:2203.15541}\ } (\bibinfo {year} {2022})}\BibitemShut {NoStop}%
\bibitem [{\citenamefont {Aidelsburger}\ \emph {et~al.}(2022)\citenamefont
  {Aidelsburger}, \citenamefont {Barbiero}, \citenamefont {Bermudez},
  \citenamefont {Chanda}, \citenamefont {Dauphin}, \citenamefont
  {{Gonz{\'a}lez-Cuadra}}, \citenamefont {Grzybowski}, \citenamefont {Hands},
  \citenamefont {Jendrzejewski}, \citenamefont {J{\"u}nemann}, \citenamefont
  {Juzeli{\=u}nas}, \citenamefont {Kasper}, \citenamefont {Piga}, \citenamefont
  {Ran}, \citenamefont {Rizzi}, \citenamefont {Sierra}, \citenamefont
  {Tagliacozzo}, \citenamefont {Tirrito}, \citenamefont {Zache}, \citenamefont
  {Zakrzewski}, \citenamefont {Zohar},\ and\ \citenamefont
  {Lewenstein}}]{aidelsburger_cold_2022}%
  \BibitemOpen
  \bibfield  {author} {\bibinfo {author} {\bibfnamefont {M.}~\bibnamefont
  {Aidelsburger}}, \bibinfo {author} {\bibfnamefont {L.}~\bibnamefont
  {Barbiero}}, \bibinfo {author} {\bibfnamefont {A.}~\bibnamefont {Bermudez}},
  \bibinfo {author} {\bibfnamefont {T.}~\bibnamefont {Chanda}}, \bibinfo
  {author} {\bibfnamefont {A.}~\bibnamefont {Dauphin}}, \bibinfo {author}
  {\bibfnamefont {D.}~\bibnamefont {{Gonz{\'a}lez-Cuadra}}}, \bibinfo {author}
  {\bibfnamefont {P.~R.}\ \bibnamefont {Grzybowski}}, \bibinfo {author}
  {\bibfnamefont {S.}~\bibnamefont {Hands}}, \bibinfo {author} {\bibfnamefont
  {F.}~\bibnamefont {Jendrzejewski}}, \bibinfo {author} {\bibfnamefont
  {J.}~\bibnamefont {J{\"u}nemann}}, \bibinfo {author} {\bibfnamefont
  {G.}~\bibnamefont {Juzeli{\=u}nas}}, \bibinfo {author} {\bibfnamefont
  {V.}~\bibnamefont {Kasper}}, \bibinfo {author} {\bibfnamefont
  {A.}~\bibnamefont {Piga}}, \bibinfo {author} {\bibfnamefont {S.-J.}\
  \bibnamefont {Ran}}, \bibinfo {author} {\bibfnamefont {M.}~\bibnamefont
  {Rizzi}}, \bibinfo {author} {\bibfnamefont {G.}~\bibnamefont {Sierra}},
  \bibinfo {author} {\bibfnamefont {L.}~\bibnamefont {Tagliacozzo}}, \bibinfo
  {author} {\bibfnamefont {E.}~\bibnamefont {Tirrito}}, \bibinfo {author}
  {\bibfnamefont {T.~V.}\ \bibnamefont {Zache}}, \bibinfo {author}
  {\bibfnamefont {J.}~\bibnamefont {Zakrzewski}}, \bibinfo {author}
  {\bibfnamefont {E.}~\bibnamefont {Zohar}},\ and\ \bibinfo {author}
  {\bibfnamefont {M.}~\bibnamefont {Lewenstein}},\ }\bibfield  {title}
  {\bibinfo {title} {Cold atoms meet lattice gauge theory},\ }\href@noop {}
  {\bibfield  {journal} {\bibinfo  {journal} {Philos. Trans. R. Soc. A}\
  }\textbf {\bibinfo {volume} {380}},\ \bibinfo {pages} {20210064} (\bibinfo
  {year} {2022})}\BibitemShut {NoStop}%
\bibitem [{\citenamefont {Dasgupta}\ and\ \citenamefont
  {Raychowdhury}(2022)}]{dasgupta_cold-atom_2022}%
  \BibitemOpen
  \bibfield  {author} {\bibinfo {author} {\bibfnamefont {R.}~\bibnamefont
  {Dasgupta}}\ and\ \bibinfo {author} {\bibfnamefont {I.}~\bibnamefont
  {Raychowdhury}},\ }\bibfield  {title} {\bibinfo {title} {Cold-atom quantum
  simulator for string and hadron dynamics in non-{{Abelian}} lattice gauge
  theory},\ }\href {https://doi.org/10.1103/PhysRevA.105.023322} {\bibfield
  {journal} {\bibinfo  {journal} {Phys. Rev. A}\ }\textbf {\bibinfo {volume}
  {105}},\ \bibinfo {pages} {023322} (\bibinfo {year} {2022})}\BibitemShut
  {NoStop}%
\bibitem [{\citenamefont {Notarnicola}\ \emph {et~al.}(2020)\citenamefont
  {Notarnicola}, \citenamefont {Collura},\ and\ \citenamefont
  {Montangero}}]{notarnicola2020real}%
  \BibitemOpen
  \bibfield  {author} {\bibinfo {author} {\bibfnamefont {S.}~\bibnamefont
  {Notarnicola}}, \bibinfo {author} {\bibfnamefont {M.}~\bibnamefont
  {Collura}},\ and\ \bibinfo {author} {\bibfnamefont {S.}~\bibnamefont
  {Montangero}},\ }\bibfield  {title} {\bibinfo {title} {Real-time-dynamics
  quantum simulation of (1+ 1)-dimensional lattice {{QED}} with rydberg
  atoms},\ }\href@noop {} {\bibfield  {journal} {\bibinfo  {journal} {Physical
  Review Research}\ }\textbf {\bibinfo {volume} {2}},\ \bibinfo {pages}
  {013288} (\bibinfo {year} {2020})}\BibitemShut {NoStop}%
\bibitem [{\citenamefont {Shaw}\ \emph {et~al.}(2020)\citenamefont {Shaw},
  \citenamefont {Lougovski}, \citenamefont {Stryker},\ and\ \citenamefont
  {Wiebe}}]{shaw2020quantum}%
  \BibitemOpen
  \bibfield  {author} {\bibinfo {author} {\bibfnamefont {A.~F.}\ \bibnamefont
  {Shaw}}, \bibinfo {author} {\bibfnamefont {P.}~\bibnamefont {Lougovski}},
  \bibinfo {author} {\bibfnamefont {J.~R.}\ \bibnamefont {Stryker}},\ and\
  \bibinfo {author} {\bibfnamefont {N.}~\bibnamefont {Wiebe}},\ }\bibfield
  {title} {\bibinfo {title} {Quantum algorithms for simulating the lattice
  {{Schwinger}} model},\ }\href@noop {} {\bibfield  {journal} {\bibinfo
  {journal} {Quantum}\ }\textbf {\bibinfo {volume} {4}},\ \bibinfo {pages}
  {306} (\bibinfo {year} {2020})}\BibitemShut {NoStop}%
\bibitem [{\citenamefont {Ciavarella}\ \emph {et~al.}(2022)\citenamefont
  {Ciavarella}, \citenamefont {Klco},\ and\ \citenamefont
  {Savage}}]{ciavarella2022some}%
  \BibitemOpen
  \bibfield  {author} {\bibinfo {author} {\bibfnamefont {A.}~\bibnamefont
  {Ciavarella}}, \bibinfo {author} {\bibfnamefont {N.}~\bibnamefont {Klco}},\
  and\ \bibinfo {author} {\bibfnamefont {M.~J.}\ \bibnamefont {Savage}},\
  }\bibfield  {title} {\bibinfo {title} {Some conceptual aspects of operator
  design for quantum simulations of non-abelian lattice gauge theories},\
  }\href@noop {} {\bibfield  {journal} {\bibinfo  {journal} {arXiv preprint
  arXiv:2203.11988}\ } (\bibinfo {year} {2022})}\BibitemShut {NoStop}%
\bibitem [{\citenamefont {Haase}\ \emph {et~al.}(2021)\citenamefont {Haase},
  \citenamefont {Dellantonio}, \citenamefont {Celi}, \citenamefont {Paulson},
  \citenamefont {Kan}, \citenamefont {Jansen},\ and\ \citenamefont
  {Muschik}}]{haase_resource_2021}%
  \BibitemOpen
  \bibfield  {author} {\bibinfo {author} {\bibfnamefont {J.~F.}\ \bibnamefont
  {Haase}}, \bibinfo {author} {\bibfnamefont {L.}~\bibnamefont {Dellantonio}},
  \bibinfo {author} {\bibfnamefont {A.}~\bibnamefont {Celi}}, \bibinfo {author}
  {\bibfnamefont {D.}~\bibnamefont {Paulson}}, \bibinfo {author} {\bibfnamefont
  {A.}~\bibnamefont {Kan}}, \bibinfo {author} {\bibfnamefont {K.}~\bibnamefont
  {Jansen}},\ and\ \bibinfo {author} {\bibfnamefont {C.~A.}\ \bibnamefont
  {Muschik}},\ }\bibfield  {title} {\bibinfo {title} {A resource efficient
  approach for quantum and classical simulations of gauge theories in particle
  physics},\ }\href {https://doi.org/10.22331/q-2021-02-04-393} {\bibfield
  {journal} {\bibinfo  {journal} {Quantum}\ }\textbf {\bibinfo {volume} {5}},\
  \bibinfo {pages} {393} (\bibinfo {year} {2021})}\BibitemShut {NoStop}%
\bibitem [{\citenamefont {Childs}\ \emph {et~al.}(2021)\citenamefont {Childs},
  \citenamefont {Su}, \citenamefont {Tran}, \citenamefont {Wiebe},\ and\
  \citenamefont {Zhu}}]{Childs2021Theory}%
  \BibitemOpen
  \bibfield  {author} {\bibinfo {author} {\bibfnamefont {A.~M.}\ \bibnamefont
  {Childs}}, \bibinfo {author} {\bibfnamefont {Y.}~\bibnamefont {Su}}, \bibinfo
  {author} {\bibfnamefont {M.~C.}\ \bibnamefont {Tran}}, \bibinfo {author}
  {\bibfnamefont {N.}~\bibnamefont {Wiebe}},\ and\ \bibinfo {author}
  {\bibfnamefont {S.}~\bibnamefont {Zhu}},\ }\bibfield  {title} {\bibinfo
  {title} {Theory of trotter error with commutator scaling},\ }\href@noop {}
  {\bibfield  {journal} {\bibinfo  {journal} {Phys. Rev. X}\ }\textbf {\bibinfo
  {volume} {11}},\ \bibinfo {pages} {011020} (\bibinfo {year}
  {2021})}\BibitemShut {NoStop}%
\end{thebibliography}%


\end{document}